\documentclass[11pt]{article}
\usepackage[normalem]{ulem}
\pdfoutput=1 

\usepackage{jheppub} 
\usepackage{url}
\usepackage{caption}
\usepackage{subcaption}

\usepackage[latin9]{inputenc}
\usepackage{float}
\usepackage{amsmath}
\usepackage{amssymb}
\usepackage{graphicx}
\usepackage{esint}
\usepackage{hyperref}
\usepackage{comment}
\usepackage{color}
\usepackage{microtype}
\usepackage{cleveref}
\usepackage{breakurl}
\usepackage{soul}
\newcommand{\be}{\begin{equation}}
\newcommand{\ee}{\end{equation}}
\newcommand{\ben}{\begin{displaymath}}
\newcommand{\een}{\end{displaymath}}
\newcommand{\bea}{\begin{eqnarray}}
\newcommand{\eea}{\end{eqnarray}}
\def\K{K{\"a}hler }
   \newcommand{\rf}[1]{(\ref{#1})}
\newcommand{\vp}{\varphi}

\def\be{\begin{equation}}
\def\ee{\end{equation}}
\def\bea{\begin{eqnarray}}
\def\eea{\end{eqnarray}}
\def\ba{\begin{array}}
\def\ea{\end{array}}
\def\bit{\begin{itemize}}
\def\eit{\end{itemize}}

\def\s{\chi}

\def\vp{\varphi}

 \makeatletter

\newcommand{\dd}{\mathrm{d}}

\allowdisplaybreaks

\makeatother

\makeatletter
\DeclareRobustCommand{\rcite}[1]{%
  \rcite@aux#1,\@nil{#1}%
}
\def\rcite@aux#1,#2\@nil#3{%
  \if\relax#2\relax
    Ref.~\cite{#3}%
  \else
    Refs.~\cite{#3}%
  \fi
}
\makeatother

\hypersetup{
    colorlinks = true,
    citecolor = {blue},
    linkcolor = {blue},
    urlcolor = {blue},
}

\title{\rm { \huge \bf \boldmath   Hybrid  $\alpha$-attractors, primordial black holes and gravitational wave backgrounds}}
\author[a,b]{Matteo Braglia,}
\author[c]{Andrei Linde,}
\author[c]{Renata Kallosh,}
\author[b,d]{Fabio Finelli}

\affiliation[a]{Center for Cosmology and Particle Physics, New York University, 726 Broadway, New York, NY 10003, USA}
\affiliation[b]{INAF/OAS Bologna, via Gobetti 101, I-40129 Bologna, Italy}
\affiliation[c]{Stanford Institute for Theoretical Physics and Department of Physics,\\ Stanford University, Stanford, CA 
94305, USA}
\affiliation[d]{INFN, Sezione di Bologna, via Irnerio 46, I-40126 Bologna, Italy}
\emailAdd{mb9289@nyu.edu}
\emailAdd{alinde@stanford.edu}
\emailAdd{kallosh@stanford.edu}
\emailAdd{fabio.finelli@inaf.it}

\abstract{We investigate the two-stage inflation regime in the theory of hybrid cosmological $\alpha$-attractors.  The spectrum of inflationary perturbations  is compatible with the latest Planck/BICEP/Keck Array results, thanks to the attractor properties of the model. However, at smaller scales, it may have a very high peak of controllable width and position, 
	leading to a copious production of primordial black holes (PBH) and generation of a stochastic background of gravitational waves (SGWB).}

\begin{document}

\maketitle

\parskip 6pt

\section{Introduction} 

There has recently been a growing interest in the possibility to generate large peaks in the amplitude of perturbations at the late stages of inflation. Such peaks  may lead to the formation of primordial black holes (PBHs)  and to generate a stochastic gravitational wave background (SGWB) which can be explored with future gravitational wave interferometers. 

One of the first fully  developed models to generate such  peaks in the primordial power spectrum (PPS) of the perturbations and consequently seed PBH formation was proposed in  \cite{Garcia-Bellido:1996mdl}  
in  the context of the  hybrid inflation scenario   \cite{Linde:1991km,Linde:1993cn}. There, it  was argued that under certain conditions such PBHs may provide a significant contribution to dark matter. 
Many other interesting mechanisms were proposed to produce a bump in the PPS from inflation. Among them is  the PBH production due to large isocurvature perturbations~\cite{Dolgov:1992pu},  amplification of perturbations during rapid turns of inflationary trajectory~\cite{Brown:2017osf,Fumagalli:2020adf,Palma:2020ejf,Braglia:2020eai,Braglia:2020taf,Iacconi:2021ltm,Kallosh:2022vha}, a sudden slowing down of the inflaton evolution often referred to as Ultra Slow Roll~\cite{Ivanov:1994pa,Garcia-Bellido:2017mdw,Motohashi:2017kbs,Germani:2017bcs,Ballesteros:2017fsr,Dalianis:2018frf,Ketov:2021fww,Cai:2021zsp,Balaji:2022rsy,Geller:2022nkr,Pi:2022zxs} or other types of features~\cite{Pi:2017gih,Ashoorioon:2019xqc,Inomata:2021uqj,Inomata:2021tpx,Dalianis:2021dbs,Boutivas:2022qtl,Inomata:2022yte} and creation of black holes and wormholes by colliding walls \cite{Garriga:2015fdk,Deng:2016vzb}.
 For a review of these and other mechanisms of 
 PBH formation and their cosmological implications see e.g. \cite{Khlopov:2008qy,Carr:2020gox,Carr:2021bzv}. 
 
In this paper, we will revisit the mechanism  of~\cite{Garcia-Bellido:1996mdl} in the recently proposed 
$\alpha$-attractor generalization~\cite{Kallosh:2022ggf} (see also~\cite{Pallis:2022mzh}) of the hybrid inflation scenario 
\cite{Linde:1991km,Linde:1993cn} and consider its consequences on PBH and SGWB production. 
While, due to the large uncertainties in the critical process of PBH formation, we will just 
touch upon it, we will instead provide a full characterization of the SGWB produced by the large 
perturbations at horizon re-entry during the radiation dominated era~\cite{Carbone:2004iv,Ananda:2006af,Baumann:2007zm} (see e.g.~\cite{Domenech:2021ztg} for a review), 
which can be used in the future to test this model. We will pay particular attention to the consistency with large scale 
measurements of the Cosmic microwave Background (CMB) anisotropies from the latest Planck/BICEP/Keck Array 
release~\cite{Planck:2018jri,BICEPKeck:2021gln,Paoletti:2022anb}, which is not a trivial task since models producing a peak in the small scale power spectrum often tend to predict a spectral index which is slightly redder than the Planck best fit. 

Let us start with a quick reminder of the basic  hybrid inflation model \cite{Linde:1991km,Linde:1993cn}. The
effective potential of this model is given by
\begin{equation}\label{hybrid}
V(\chi,\phi) =  {1\over 4\lambda}(M^2-\lambda\chi^2)^2 + {m^2\over 2}\phi^2 + {g^2\over 2}\phi^2\chi^2\ .
\end{equation}
For $\phi > \phi_c = M/g$ the only minimum of the effective
potential
$V(\chi,\phi)$ with respect to $\chi$, sometimes referred to as {\em hybrid field},  is at $\chi = 0$. The  
mass squared of the  field $\chi$ at $\chi = 0$ is equal to
$V_{ \chi,\chi}(\chi = 0) = -M^{2}+g^{2}\phi^{2}.$ At large $\phi$, the curvature of the effective
potential in the $\chi$-direction is much larger than in the 
$\phi$-direction. Thus, during the first stages of
expansion of the universe, the field $\chi$ rolls down to $\chi =
0$, whereas the field $\phi$ could remain large and drive inflation for a much longer
time.    At that time, the potential of the inflaton field $\phi$ is given by $V(\phi) = {m^2\over
2}\phi^2 +V_{\rm uplift}$, where the uplift potential is given by ${M^{2}\over 4\lambda}$.

At the moment when the inflaton field $\phi$ becomes smaller than  $\phi_c = M/g$, the  effective
mass squared of the field $\chi$ at $\chi = 0$ becomes negative (tachyonic),  quantum fluctuations of this field begin to grow, and a so-called {\em waterfall} phase transition
with symmetry breaking occurs. 

In the simplest versions of the hybrid inflation scenario, the absolute value of the tachyonic mass of the field $\chi$ becomes much greater than the Hubble constant $H$  shortly after the field $\phi$ becomes smaller than   $\phi_{c}$.  In that case, the tachyonic instability leads to an abrupt end of inflation at  $\phi  \approx \phi_{c}$   \cite{Linde:1991km,Linde:1993cn}. 
However, in some models the tachyonic mass of the field $\chi$ at $\phi  \lesssim \phi_{c}$ may remain  much smaller  than the Hubble constant $H$. In the  Higgs-type potential $V(\chi)\sim (\chi^{2}- \chi_{0}^{2})^{2}$ used  in  \cite{Linde:1991km,Linde:1993cn,Garcia-Bellido:1996mdl}, this regime occurs for  $\chi_{0} \gg 1$ (in Planck mass units  $M_{\rm pl}=1$).  In such cases, inflation continues while the field $\chi$ slowly rolls down  towards the minimum of $V(\chi)$. As we will explain later, the amplitude of perturbations produced at the second stage of inflation in  hybrid inflation models  with such properties can be extremely large.  

In fact, these perturbations can be so large that in addition to producing PBH they may also trigger the process of eternal inflation inside some regions of the observable part of the universe \cite{Garcia-Bellido:1996mdl}. One way to avoid this problem and regularize the amplitude of the perturbations is to consider models with $\chi_{0} \lesssim 1$ \cite{Garcia-Bellido:1996mdl}. However,  this does not  solve the problem of superheavy topological defects in this scenario  and in its various generalizations, such as the Clesse-Bellido model \cite{Clesse:2015wea}.

In this paper  we are going to show that both of these problems can be solved by  adding a tiny linear term $\sim \chi$ to the potential~\eqref{hybrid}. As we will see, this allows to control the strength of the effect and the amplitude of the peak in the PPS, and simultaneously  solve the problem of topological defects.
 
 Whereas the basic mechanism of the PBH production described above~\cite{Garcia-Bellido:1996mdl} is quite general and can be implemented in a broad class of hybrid inflation models, the simplest versions of hybrid inflation \cite{Linde:1991km,Linde:1993cn} resulted in the spectral index $n_{s} \approx 1+ 2m^{2}/V_{\rm uplift} > 1$, which did  not match the Planck data \cite{Planck:2013jfk}.  

An attempt to overcome this problem was made in  \cite{Clesse:2015wea}. The authors  changed the sign of $m^{2}$ making the inflaton potential tachyonic, $m^{2} < 0$, which allowed to have $n_{s} < 1$. This paper, together with the detection of gravitational waves from a binary black hole merger by the LIGO-Virgo collaboration~\cite{Abbott:2016blz},  attracted new attention to the possibility pointed out in~\cite{Garcia-Bellido:1996mdl} that PBH may constitute a substantial portion of the dark matter~\cite{Bird:2016dcv,Sasaki:2016jop,Clesse:2016vqa}.  However,   the tachyonic potential $V(\phi)$ used in the toy model discussed in   \cite{Clesse:2015wea} is unbounded from below, and the field $\phi \gtrsim \phi_{c}$ instead of moving towards $\phi = 0$ tends to run  to $\phi \to\infty$. This makes it difficult to solve the problem of initial conditions for inflation in this scenario unless the potential is further modified at $\phi>\phi_c$.

Fortunately, this problem can be addressed in the $\alpha$-attractor versions \cite{Kallosh:2022ggf}  of the  original hybrid inflation scenario \cite{Linde:1991km,Linde:1993cn}. For a broad choice of parameters, these models have the standard universal $\alpha$-attractor predictions  \cite{Kallosh:2013hoa,Ferrara:2013rsa,Kallosh:2013yoa,Kallosh:2021mnu,Kallosh:2022feu} matching the Planck/BICEP/Keck Array data   \cite{Planck:2018jri,BICEPKeck:2021gln}. In particular,   $n_{s }= 1-  {2  \over  N} \sim 0.96 - 0.97$ for the exponential attractors, where $N$ is the number of $e$-foldings \cite{Kallosh:2013yoa}. However,  in the large  uplift limit $V_{\rm uplift} \gg {m^2\over
2}\phi^2$  these models have another attractor regime, with $ n_s = 1$. By changing the parameters, one can continuously interpolate between the two attractor predictions, $n_{s }= 1-  {2  \over  N}$ and $n_{s}  = 1$  \cite{Kallosh:2022ggf}. Because of the double attractor structure of such models,  their predictions are compatible with all presently available cosmological data.  In the models where the tachyonic mass of the field $\chi$ at $\phi  \lesssim \phi_{c}$  is    smaller  than the Hubble constant $H$, which we will study in this paper, the range of possible values of $n_{s}$ is even broader.  

In this paper,  we will study the simplest versions of such models,   explain how one can solve the problem of initial conditions there, and  evaluate the spectrum of perturbations in these models. 
We will demonstrate that one can keep $n_{s}$ and $r$ consistent 
with the latest CMB data, while simultaneously  controlling the shape, 
the width and the height of the peak of the spectrum of perturbations 
which determine the features of the PBHs and the  stochastic SGWB background  produced in this scenario.

This paper is organized as follows. We start with a discussion of hybrid inflation in the context of $\alpha$-attractors in Section~\ref{sec:model}.   After reviewing the basic results of~\cite{Kallosh:2022ggf}, we discuss the problem of initial conditions in our model in Section~\ref{sec:initial}. In Section \ref{sec:single}, we will make a step back and study the perturbations in the $\chi$-sector of hybrid inflation, temporarily ignoring the field $\vp$. As we will see, this will help to understand the origin of the high peak of perturbations produced   in various versions of the hybrid inflation scenario.  In Section~ \ref{sec:main}, we explain in details how the model parameters control distinct properties of the bump in the PPS and discuss the constraints from Planck/BICEP/Keck Array data.  Then, in Section~\ref{sec:SGWB}, we describe the gravitational wave phenomenology of our model providing clear targets for gravitational wave interferometers and discussing how to interpret a future detection of a SGWB in terms of our model parameters. In Section~\ref{sec:polynomial} we will generalize our results by considering models based on polynomial attractors or KKLTI potentials~\cite{Kallosh:2022feu}. We conclude in Section~\ref{sec:conclusions}. In the Appendix \ref{sec:sugra}, we provide an implementation of our model in SUGRA constructions.  We set $M_{\rm pl}=1$ throughout our analysis. The numerical results in this paper are obtained using a multifield extension of the BINGO code~\cite{Hazra:2012yn}, presented in~\cite{Braglia:2020fms}, which is however not yet public.

\section{Hybrid   $\alpha$-attractors}
\label{sec:model}

Before describing the hybrid attractors~\cite{Kallosh:2022ggf}, we will represent the original model~\rf{hybrid}, together with a small additional linear term $\mu^{3} \chi$, in a slightly different form, which is more convenient for our investigation\footnote{One should also add a tiny constant term $\Delta V \approx -\mu^{3} \chi_{0}$ to the potential~\eqref{h} to ensure nearly vanishing value of the cosmological constant, but it can be ignored in the discussion of inflation.}:
 \be\label{h}
V(\chi,\phi) =M^{2}\left[{({\chi^2 -  \chi_{0}^{2}})^2\over 4\chi_{0}^{2}} 
+ {{\tilde m}^2\over
2 }\phi^2 + {{\tilde g}^2\over 2 }\phi^2\chi^2 +d\, \chi  \right].
\ee
Here   $\tilde m = m/ M$,     $ \tilde g = g/M$, and $d =  {\mu^{3}\over M^{2 }}$.  In this paper we will consider models with $\chi_{0} = O(1)$ and $\mu  \sim M \sim m \sim    g \sim 10^{{-5}}$, with masses given in the Planck mass units $M_{\rm pl}=1$.  In that case $\tilde m \sim \tilde g  = O(1)$, whereas the linear term can be very small,  with $d \sim10^{{-5}}$. Therefore it  does not affect early stages of inflation. However, as we will see, it plays an  important role in the theory of the PBH production.

 Note that the  linear term slightly breaks the symmetry $\chi \to -\chi$ of  the original hybrid inflation scenario. This symmetry was introduced in  \cite{Linde:1991km,Linde:1993cn} for simplicity, it is not a fundamental requirement, and in our context there is no gauge symmetry protecting it. Thus one may argue that it is technically natural.  Its possible interpretation in the context of supergravity is discussed in  the Appendix \ref{sec:sugra}.

In the original version of the hybrid inflation  \cite{Linde:1991km,Linde:1993cn},  both $\phi$ and $\chi$ are canonically normalized. To present the simplest $\alpha$-attractor generalization of the original model, it is sufficient  to modify the kinetic term of the  field $\phi$:
  \be 
{ {\cal L} \over \sqrt{-g}} =  {R\over 2}  -  {(\partial_{\mu} \phi)^2\over 2\bigl(1-{\phi^{2}\over 6\alpha}\bigr)^{2}} -  {(\partial_{\mu} \chi)^2\over 2} - V(\chi,\phi)   \,  .
\label{cosmoAA2}\ee 
One may also make a similar generalization of the kinetic term of the field $\chi$    \cite{Kallosh:2022ggf}, but  it is not required  in the context of our investigation. The geometric interpretation of this modification can be found in~\cite{Kallosh:2015zsa} and its supergravity implementation for hybrid attractors is given in~\cite{Kallosh:2022ggf} and in the Appendix~\ref{sec:sugra} of our paper.

Upon  a transformation to the canonical variable $\vp$, the hybrid inflation potential becomes
 \bea\label{hybridab}
 V(\chi,\vp)=M^{2}\left[{({\chi^2 -  \chi_{0}^{2}})^2\over 4\chi_{0}^{2}} 
  +3\alpha (\tilde m^{2 } + \tilde g^2 \chi^{2}) \tanh^{2}{\varphi\over\sqrt {6 \alpha}}  +d\, \chi \right].
\eea
The potential at $\chi = 0$ is given by
\be 
V(\vp)=     M^{2}\left({\chi_{0}^{2}\over 4}+  3{\tilde m}^{2 }\alpha \tanh^{2}{\varphi\over\sqrt {6 \alpha}}\right)  \ .
\ee
This is the simplest $\alpha$-attractor potential  \cite{Kallosh:2013hoa,Ferrara:2013rsa,Kallosh:2013yoa,Kallosh:2021mnu,Kallosh:2022feu}, uplifted by the term $V_{\rm uplift}={M^{2}\chi_{0}^{2}\over 4}$. Another important parameter is the mass squared of the field $\chi$ at $\chi = 0$,
\be
M^{2}_{\chi} = V_{ \chi,\chi}(\chi = 0) =  M^{2}\left(-1+6\alpha \tilde g^{2}\tanh^{2}{\varphi\over\sqrt {6 \alpha}}\right) . 
\ee
 The last term in this equation stabilizes the inflationary trajectory $\chi = 0$  at large  $\vp$. 
 
 In this paper, we will consider the case $6\alpha \tilde g^{2} >1$. In that case $M^{2}_{\chi} > 0$ at sufficiently large $\vp >\vp_{c}$, where 
\be\label{pc}
\tanh^{2}{\varphi_{c}\over\sqrt {6 \alpha}} =  {1\over 6\alpha \tilde g^{2}}.
\ee
 During inflation, when the field $\vp$ decreases below   $\vp_{c}$, the mass squared of the field $\chi$ becomes negative and the tachyonic instability with generation of the scalar field $\chi$ develops. At $\vp = 0$ this mass has its largest absolute value, $M^{2}_{\chi}(\vp = 0) =  -M^{2}$.
 
\begin{figure}
\centering
\includegraphics[scale=0.42]{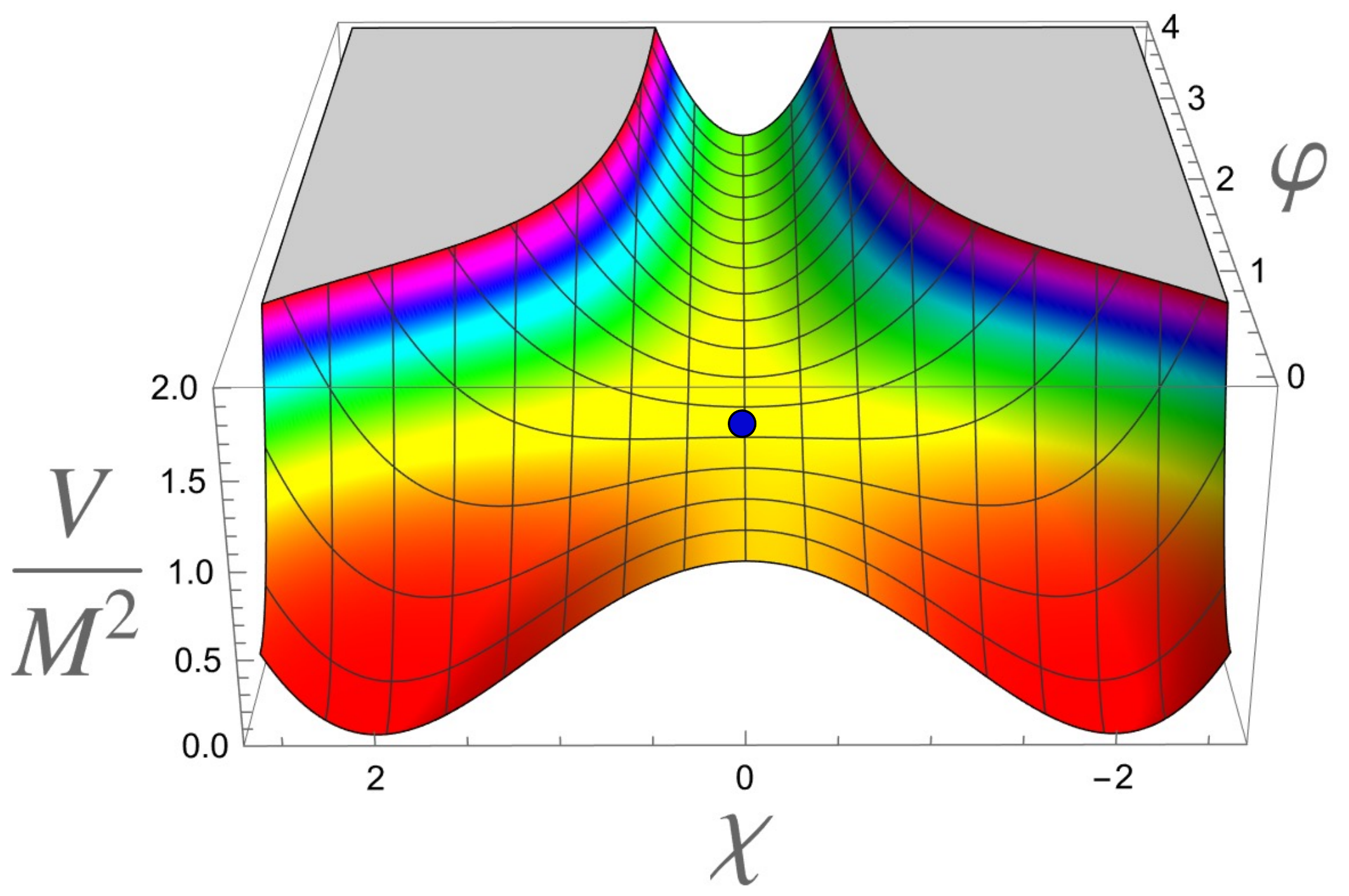}
\caption{\footnotesize  Hybrid inflation potential for the model \rf{hybridab} with $\tilde{m} = 1/3,   \chi_{0} =  2, \tilde{g} = 1, \alpha = 1,\,d=-5\times10^{-6}$. The blue dot corresponds to $\vp_{c}$ \rf{pc}, below which the mass of the field $\chi$ becomes tachyonic.  Note that the effects of the linear term are almost invisible by eye.}  
\label{h2}
\end{figure}

 The potential \rf{hybridab}  for some particular values of parameters  is shown in Fig. \ref{h2}.

It is instructive to compare  $M^{2}_{\chi}(\vp = 0) =  -M^{2}$ with the Hubble constant at $\vp = 0$,
\be
H_{0} = \sqrt{V_{\rm uplift}\over 3} = {M  \chi_{0}\over 2\sqrt 3 } \ .
\ee  
The main regime explored in the   original formulation of the hybrid inflation scenario was $\chi_{0} \ll1$  \cite{Linde:1991km,Linde:1993cn}, in which case the absolute value of the tachyonic mass $M_{\chi}$ becomes much larger than the Hubble constant as soon as $\phi$ becomes smaller than $\phi_{c}$. This leads to an abrupt termination of inflation  at $\phi \approx  \phi_{c}$  \cite{Linde:1991km,Linde:1993cn,Kallosh:2022ggf}.

In this paper, following  \cite{Garcia-Bellido:1996mdl}, we will be interested in the opposite regime $\chi_{0} \gtrsim 2\sqrt 3$. In this regime  $|M_{\chi}(\chi = 0)| \lesssim H$ for all $\vp < \s\lesssim \vp_{c}$. This mass vanishes when the field $\varphi$ is close to $\vp_{c}$. As a result, the tachyonic instability is very slow to develop. Importantly, at any nonzero value of $\vp$, its contribution to the equation of motion of the field $\chi$ is only slowing it down, and at $\vp \ll  \vp_{c}$ the contribution of the field $\vp$ to the equation of motion of the field $\chi$ becomes negligible. 
  
  That is why one can learn quite a lot about the waterfall regime in such models by studying  the  evolution of the field $\chi$ from the top of  the single-field inflationary potential $V(\chi) =  ({\chi^2 -  \chi_{0}^{2}})^2\,M^2\,/\,4\chi_{0}^{2}$, ignoring the field $\vp$. We will discuss it in Section \ref{sec:single}. This will explain in an intuitive way why it is natural to expect a very high peak for the inflationary perturbations in such models, and how one can control its height. In subsequent sections, we will return to the full scenario describing classical and quantum evolution of both fields. 
 
 But before doing it, we will discuss the problem of initial conditions in this scenario. And the first question is: How did the scalar fields found their way into the narrow infinitely long valley with $\chi = 0$?
 
  \section{Initial conditions for inflation}\label{sec:initial} 

As we can see from equation \rf{h},  the potential  $V(\chi,\vp)$ at $\vp \gg \sqrt{6\alpha}$ 
practically does not depend on $\vp$ because in this regime the function $\tanh {\varphi_{c}\over\sqrt {6 \alpha}}$ 
becomes exponentially close to $1$:   
\bea\label{hybridab2}
 V(\chi,\vp)\approx M^{2}\left[{({\chi^2 -  \chi_{0}^{2}})^2\over 4\chi_{0}^{2}} 
  +3\alpha (\tilde m^{2 } + \tilde g^2 \chi^{2}) +d \,\chi  \right].
\eea
Thus at large $\vp$ one can apply to this theory the standard arguments of the single field 
chaotic inflation scenario \cite{Linde:1983gd,Linde:1985ub,Linde:2017pwt}. One can consider 
a tiny Planck size domain of the universe with the Planckian value of the potential energy 
of the field $\chi$, and let it fall. If the kinetic and gradient energy of the fields initially 
are smaller that its potential energy, each such tiny domain becomes exponentially expanding. 

The potential $ V(\chi,\vp)$ approaches the Planck boundary at $\chi^{2} \sim 2\chi_{0}/M \gg 1$ 
for any value of $\vp$ in the infinitely large  range $-\infty < \vp < +\infty$. Therefore in this 
context it is natural to consider initial conditions where both fields are extremely large, but 
$\vp \gg \chi$. For most of such initial conditions the field $\chi$  gradually falls down to the 
valley, oscillates with a rapidly decreasing amplitude, and relaxes at the bottom of the valley, 
and, only after that, the  field $\vp$ begins to move towards $\vp = 0$.   

\begin{figure}
	\begin{center}
		\includegraphics[width=.5\columnwidth]{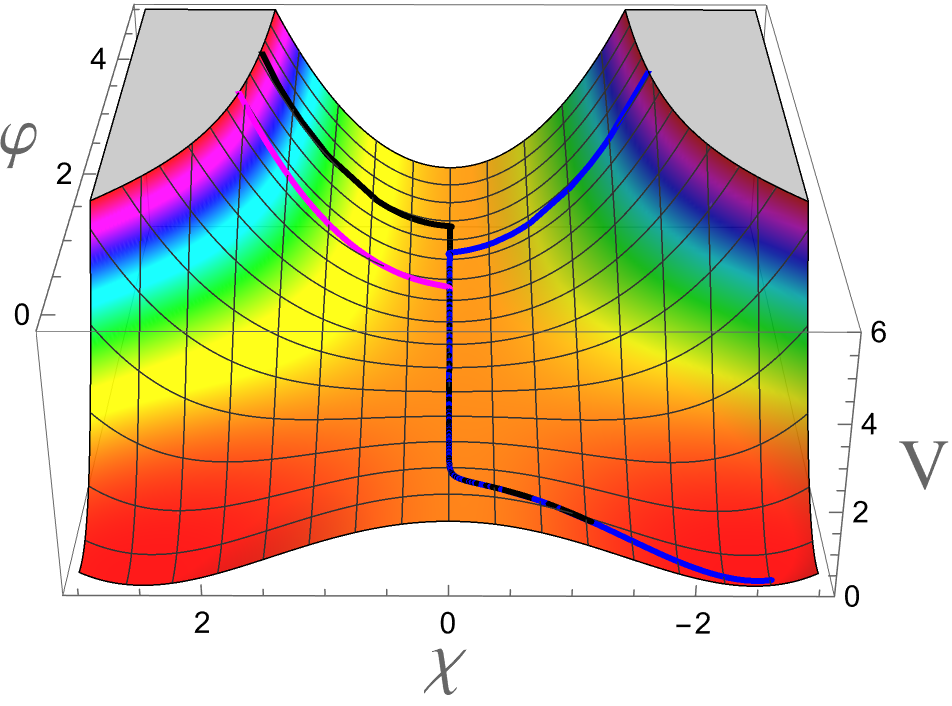} 
		\includegraphics[width=.49\columnwidth]{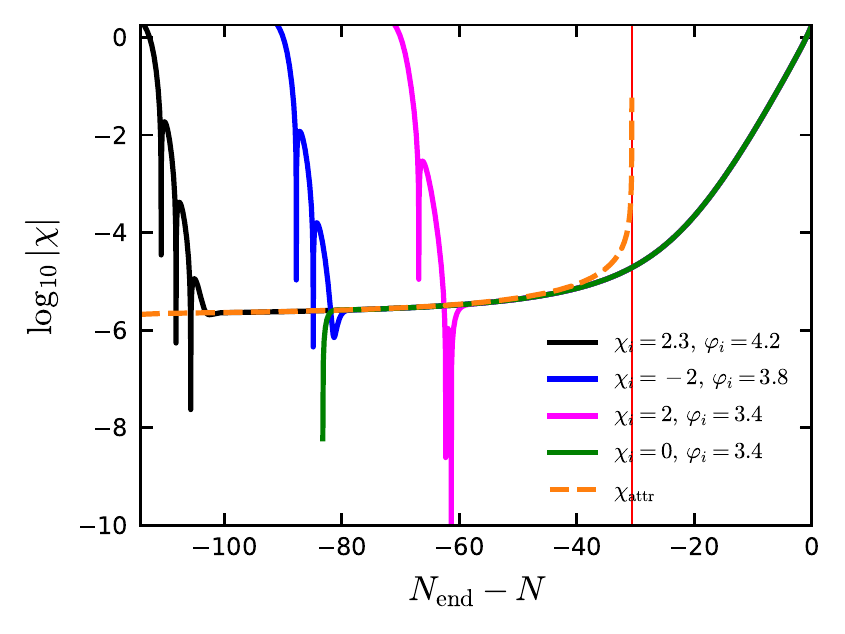}
	\end{center}
	\vskip -15pt
	\caption{\footnotesize	[Left] 
		Initial conditions for hybrid attractors. 
		The parameters used are those in Section~\ref{sec:main},~ Eq.~\rf{eq:baseline_parameters}. 
		 The potential is shown in units of $M^{2}$. 
		The fields begin at large values of $\vp$ and $\chi$. Then $\chi$ rapidly falls down, and, 
		after that, inflation driven by the  field $\varphi$ begins. Eventually, after $\varphi$ 
		crosses $\varphi_c$ the field $\chi$ falls to one of the two minima of the potential  
		$\chi \approx \pm \chi_{0}$  depending on the sign of $d$.   The initial conditions are shown on the right panel.  
		[Right] Evolution of the  field $\chi$ for different initial conditions compared to 
		the attractor solution in Eq.~\eqref{min}.  The agreement of the attractor solution 
		with the numerical (exact) solution of $\chi$ is remarkable. As expected, it starts 
		to deviate from the position of the  minimum of the potential \rf{min} close to when 
		$\vp\sim\vp_c$. The power spectrum is clearly unaffected by the initial conditions.}
	\label{fig:init} 
\end{figure}

Note that, because of the tiny linear term $d\, \chi$, the field $\chi$ does not relax at $\chi = 0$, 
but it is slightly displaced in the direction depending on the sign of the coefficient $d$. 
At $\chi \ll \chi_{0}$ and  $d \not = 0$ the valley is at the minimum of the potential 
\be\label{hybridab3}
V(\chi,\vp) \approx  {M^{2} }\left[{1\over 2}\left(-1+6\alpha \tilde g^{2}
\tanh^{2}{\varphi\over\sqrt {6 \alpha}}\right) \chi^{2} + d\, \chi \right] \ .
\ee
 This potential admits an attractor for $\chi$ defined by its minimum at 
\be\label{min}
\chi_{\rm attr}=  -{ d \over  6\alpha \tilde g^{2}\tanh^{2}{\varphi\over\sqrt {6 \alpha}}-1}  \ .
\ee 
As mentioned in the previous section, we study models with $6\alpha \tilde g^{2}> 1$. 
At  $\vp \gg \sqrt {6 \alpha}$, we have $\tanh^{2} \varphi/\sqrt {6 \alpha}\simeq1$ 
and the minimum with respect to $\chi$ is at a constant $\vp$-independent distance from $\chi = 0$:
\be\label{equilibrium}  
\chi =  -{ d \over  6\alpha \tilde g^{2}-1}  \ .
\ee
Because of the rapid decrease of the kinetic energy of the field $\chi$ during inflation, the amplitude of its oscillations about this minimum decreases exponentially fast. Then, the field $\vp$ starts slowly moving towards  its smaller values, see Fig. \ref{fig:init}. 
At $ \sqrt {6 \alpha} \gtrsim \vp > \vp_{c}$, the minimum of this potential moves further away from zero.
Close to $\vp = \vp_{c}$, the part of the potential $\sim \chi^{2}$ 
vanishes, so it can no longer protect the field $\chi$ from growing  due to the linear term $d\,\chi$. 

Thus, even though this term $d\,\chi$ is very small, it pushes the field $\chi$  away from the $\chi = 0$. Eventually, the fields falls down to one of the two minima at $|\chi| \approx \chi_{0}$ at $\vp = 0$.  The choice of the minimum is completely independent on the initial conditions on the two fields, and rather depends on the sign of the coefficient $d$ of the tiny linear term.  

A more important statement is that if the initial values of the field $\vp$ are sufficiently large, the original oscillations of the field $\chi$ becomes exponentially damped, and the final results of  our  calculations of the spectrum of perturbations do not depend on initial conditions.

 Simplicity and robustness of this process is one of the advantages of the original hybrid inflation scenario, as well as of its $\alpha$-attractor generalization. Its most important part is the existence of a long valley to which the field $\chi$ falls, and the positive slope of the potential with respect to the field $\vp$ which pushes it along the valley towards $\vp = 0$  \cite{Linde:1991km,Felder:1999pv,Clesse:2008pf}.

\begin{figure}
	\begin{center}
		\includegraphics[width=.5\columnwidth]{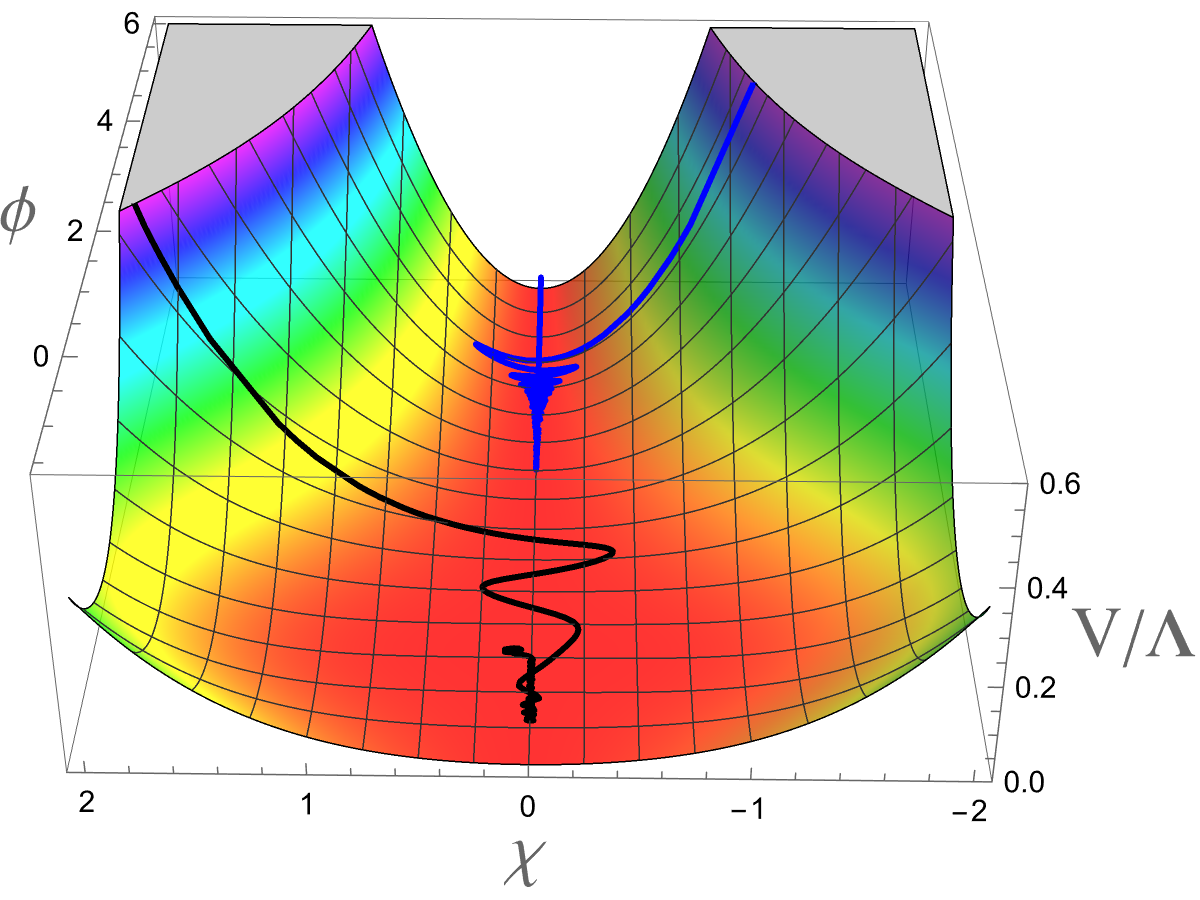}
	\end{center}
	\caption{\footnotesize	 Initial conditions problem for the model (1) of Ref. \cite{Clesse:2015wea}.  If evolution begins at smaller values of $\phi$, as shown by the black line, the field evolution is chaotic, and hybrid inflation regime typically does not emerge. If evolution begins at greater values of $\phi$, as shown  by the  blue line, the field $\phi$ turns, runs to infinity, its tachyonic potential becomes negative,  and the universe collapses.  In this example, we use the parameters of Fig.~3 in Ref.~\cite{Clesse:2015wea}   with $\phi_c=0.1$ and $(\phi_i,\,\chi_i)\,=(3,\,3),\,(4,\,-2)$ for the black and blue lines respectively. }
	\label{fig:init2} 
\end{figure}

The situation can be very different if one attempts to  make the parameter $m^{2}$ negative, as 
in the  toy model described in \cite{Clesse:2015wea}. In that case, if one considers small initial values of the field $\phi$, the field moves to $\phi = 0$, but without a significant fine-tuning of initial conditions it is difficult to achieve a long hybrid inflation  regime starting with large $\chi$, see Fig. \ref{fig:init2}. On the other hand, if one considers large initial values of $\phi$, the field $\phi$ initially moves to smaller $\phi$, but then it turns and rolls down along the valley towards very large $\phi$ with $V(\phi)< 0$, and the universe collapses.  It is possible to solve this problem by a proper modification  of  the potential used in \cite{Clesse:2015wea} at $\phi > \phi_{c}$.   Fortunately, this issue does not appear in the original version of the hybrid inflation scenario  \cite{Linde:1991km,Linde:1993cn},  and in the hybrid attractor models  described in our paper.

 \section{Evolution of the field $\chi$ in the single-field approximation}\label{sec:single} 
As we have argued in  section \ref{sec:model}, one can learn quite a lot about the waterfall stage of inflation in hybrid inflation with $\chi_{0} \gtrsim 2\sqrt 3$ by ignoring the field $\vp$ and investigating the single field model with
\be\label{hill2}
V(\chi) = {M^{2}\over 4\chi_{0}^{2}}  ({\chi^2 -  \chi_{0}^{2}})^2 \ .
\ee
Many of our results will be valid for any single field inflationary potential; they will be based on the general theory of eternal inflation \cite{Steinhardt:1982kg,Linde:1982ur,Vilenkin:1983xq,Linde:1986fd} which we will remind here following the general theory of eternal inflation developed in \cite{Linde:1986fd}.

In the slow-roll approximation, ignoring quantum fluctuations, during each $e$-folding the scalar field decreases by 
\be\label{quant}
\Delta \chi = {V_{\chi}\over V} \ .
\ee
where $V_{\chi }= {\partial V\over \partial \chi}$. During that time the volume of each inflationary domain of the horizon size $\mathcal{O}(H^{-1})$  increases $e^{3} \sim 20$ times, so we get $\sim 20$ independent horizon-size inflationary domains where the average value of the field $\chi$ decreases by $\Delta \chi = {V_{\chi}/ V}$. 

However, during the same time, inflationary quantum fluctuations may increase the value of $\chi$ by 
\be 
\delta \chi \sim {H\over 2\pi} = {\sqrt V\over 2\sqrt 3\,  \pi} \ .
\ee 
If $\delta \chi  > |\Delta \chi|$, these quantum jumps may bring half of the 20 horizon-size domain uphill, back to where we started. This leads  to the regime of eternal inflation \cite{Linde:1986fd}. Remarkably, as one can easily see, the condition $\delta \chi  < |\Delta \chi|$ required for the absence of eternal inflation is equivalent to the condition that the amplitude of perturbations is smaller than $\mathcal{O}(1)$: 
\be\label{as}
A_{s} ={V^{3}    \over 12 \pi^{2} V_{\chi}^{2}} \lesssim 1.
\ee
This is not a coincidence, since eternal inflation would imply that inflation continues in some parts of the universe, whereas in many other parts of the universe inflation is over and the energy density rapidly becomes small.

In the context of the model \rf{hill2}, the criterion \rf{as} is always violated at the top of the potential where $V_{\chi}\sim0$. However, quantum fluctuations in each horizon size domain during one $e$-folding of inflation shift the field $\chi$ from $\chi = 0$ by $\delta \chi \sim {H\over 2\pi}$. Thus one can use this estimate for the initial beginning of the inflationary trajectory. Expanding $V(\chi) = V  +  V_{\chi,\chi} \chi^{2}/2$ we find a simple  criterion for the existence of eternal inflation at the top of the potential: $|V_{\chi,\chi}| \lesssim  V$. 

A similar and perhaps a slightly more accurate estimate can be obtained by investigation of  eternally  inflating domain walls separating parts of the universe with $\chi > 0$ from the parts with $\chi < 0$ \cite{Linde:1994hy,Vilenkin:1994pv,Linde:1994wt}. According to \cite{Sakai:1995nh}, this happens for 
\be\label{sakai}
{|V_{\chi,\chi}|\over V} < {9\over 2\pi} \ .
\ee 
This means that if this slow-roll condition is satisfied at the top of the potential, inflation there is eternal.

In application to the model \rf{hill2},  eternally inflating topological domain walls appear if 
\be\label{s2}
 M \lesssim 2 H \ , 
\ee
or, equivalently, if \cite{Sakai:1995nh}
\be
 \chi_{0} \gtrsim  1.7 \  .
\ee

What relation does it have to the two-field  evolution in the hybrid inflation scenario? The answer is that if the field $\vp$ initially was large, during the long process of its rolling along the valley of the potential at $\vp > \vp_{c}$ any original oscillations of the field $\chi$ disappear, and it becomes stabilized at $\chi = 0$. Then when the field $\vp$ passes the critical point $\vp_{c}$, the field $\chi$ may start falling down. Since $V_{\chi} = 0$ at $\chi = 0$, the field $\chi$ can fall down from $\chi = 0$ only due to generation of inflationary perturbations.

Note that while the field $\vp$ rolls to $\vp= 0$, the Hubble constant decreases, whereas the absolute value of the tachyonic mass squared $|M^{2}_{\chi}|$ increases until  it reaches $M^{2}$. Therefore if the  condition \rf{sakai}, \rf{s2}  is  satisfied at $\vp = 0$, it is  satisfied everywhere along the ridge $\chi = 0$, $\vp < \vp_{c}$. In that case we enter the eternal inflation regime, the process which generates enormous density perturbations on the scale well within the observable part of the universe. There perturbations are sufficient for producing not only PBH, but also eternally  inflating domain walls separating parts of the universe with $\chi > 0$ from the parts with $\chi < 0$.

Thus the first conclusion which follows from this analysis is that in hybrid inflation with $M \lesssim H$ it is very easy to produce PBH. The only problem here is how to  avoid overproducing PBH and inflating topological defects.

The simplest way to avoid eternal inflation and to decrease the amplitude of perturbations below $O(1)$ is to consider models with $\chi_{0}\ll1$. It is still possible to have a second stage of fast roll inflation in this scenario,  producing large perturbations~\cite{Garcia-Bellido:1996mdl,Linde:2001ae,Clesse:2015wea}.  However, this does not address the problem of superheavy topological defects in such models. 

To solve both problems simultaneously, we added the small term $\mu^{3}\, \chi$ to the potential  \rf{h}, see \rf{hybridab2}. In that case one has $V_{\chi}(\chi = 0,\vp= 0) = \mu^{3}$, and the condition $A_{s } < 1$ reads
\be \label{cc}
\mu^{3} \gtrsim {V^{3/2}    \over  2\sqrt 3 \pi }    \ .
\ee
To give a particular example, for  $V \sim 10^{{-10}}$  this condition implies that $\mu^{3} \gtrsim 10^{-16}$.  This means that by adding a tiny term $10^{-16} \chi$ to the potential one can avoid the problems discussed above.
The corresponding constraint on $d =  \mu^{3}/M^{2}$ is  
\be\label{dd}
d \gtrsim {M \chi_{0}^{3}\over 2^{4}\, \sqrt 3   \pi \, } \ .
\ee
As we mentioned in section \ref{sec:initial}, the term $ d\,\chi $ leads to an additional modification of this scenario. In its absence, the field $\chi$ is rolling along the minimum of the potential at $\chi = 0$, which is why in our previous estimates of the amplitude of the perturbations we used $\chi = 0$ as a starting point of the inflationary waterfall. However, the term $d\,\chi $ pushes the field $\chi$ slightly away from $\chi = 0$. This happens well before $\vp$ becomes smaller than $\vp_{c}$, see  \rf{equilibrium}, \rf{min}.
Thus, even though the linear  term $d\,\chi$ is very small, it pushes the field $\chi$  away from the ridge of the potential and from its initial equilibrium state \rf{min}.  This additionally decreases the height of the peak of the perturbations. On the other hand, during the several e-foldings near $\vp_{c}$ the average amplitude of the perturbations may grow slightly above $H\over 2\pi$, by a factor $O(1)$. Therefore it would be interesting to perform a more detailed investigation of stochastic effects during inflation, following   \cite{Starobinsky:1986fx,Goncharov:1987ir,Linde:1993xx,Garcia-Bellido:1993fsr,Felder:2000hj,Felder:2001kt,Finelli:2008zg,Finelli:2010sh,Demozzi:2010aj,Kawasaki:2015ppx,Assadullahi:2016gkk,Pinol:2020cdp}.  However, we believe that the simple estimates \rf{cc}, \rf{dd} give a good estimate of the range of validity of the  perturbative analysis to be used  in this paper. 

Thus we see that by adding the term $d \chi$ to the potential, one can control the height of the peak of the perturbations. 
A detailed  analysis of the perturbations produced in the full two-field scenario is rather sophisticated and  will be given in the next section.  Interestingly, we will find that the spectrum of the perturbations produced at $\vp$ sufficiently far below the critical point $\vp_{c}$ is well described by the theory of perturbations in the single-field model described in this section, while around the critical point $\varphi_c$ multifield effects play a non-negligible role.

\section{Hybrid exponential attractors and the spectrum of perturbations}
\label{sec:main}
\begin{figure}
 \vskip -10pt
\centering
\includegraphics[scale=0.35]{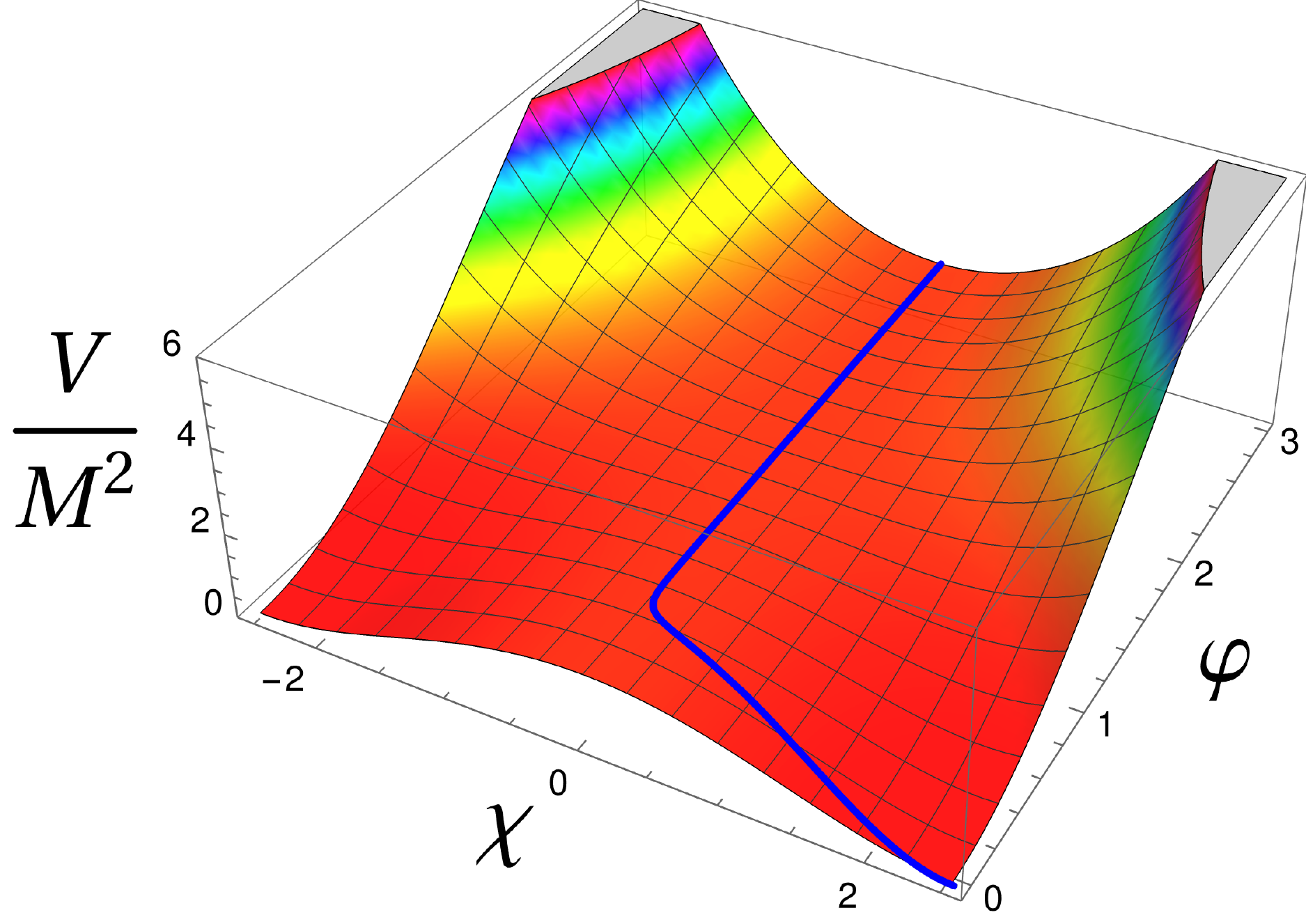} \hskip 20 pt 
\includegraphics[scale=0.35]{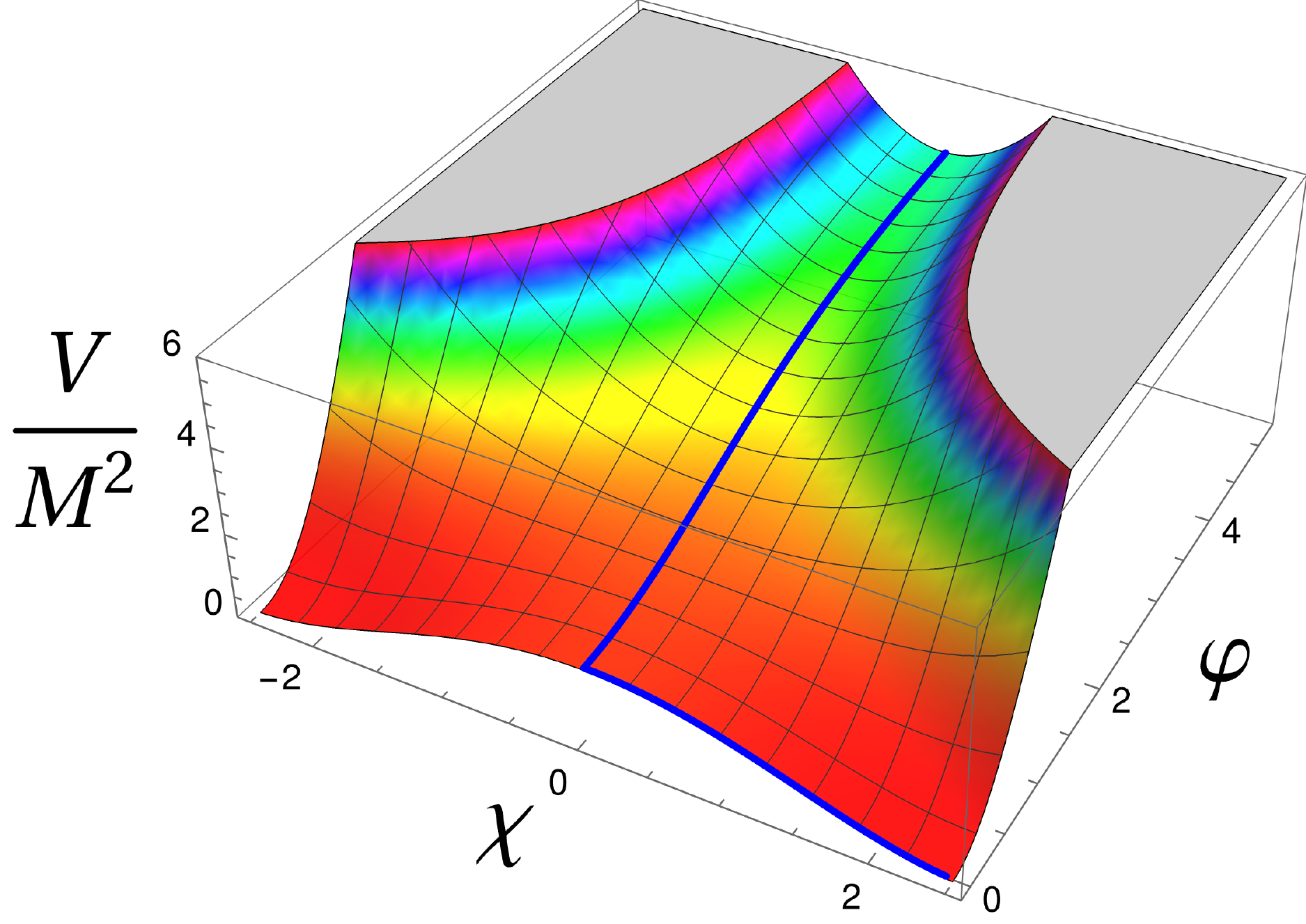}
\caption{\footnotesize  Trajectory in the exponential attractor model for the baseline parameters in Eq.~\eqref{eq:baseline_parameters}  (left) and for the parameters in Eq.~\eqref{eq:baseline2_parameters} (right). To better appreciate the differences, we set the same range for the y-axis. }
\label{fig:traj_exp} 
\end{figure}

 In this section we will perform a full numerical investigation of perturbations in hybrid attractors.  
For illustration, we will consider the exponential attractor model in Eq.~\eqref{h} and consider the following combination of parameters as a benchmark
\begin{equation} 
	\label{eq:baseline_parameters}
	M=1.47\times10^{-5}\,,\,\,\,\,\, 		\alpha=1\,,\,\,\,\,\, 		\tilde{g}=0.8\,,\,\,\,\,\, 		\tilde{m}=0.3\,,\,\,\,\,\, 		\chi_0=2.5\,,\,\,\,
d=-5\times10^{-6}\,,
\end{equation}
 which we denote as the {\em baseline} parameters and explore how changing each parameter affects the shape and amplitude of the bump  in the primordial power spectrum that is produced at small scales. Note that we always set the  field $\chi$ initially at rest at $\chi_i=0$ and we adopt the initial condition $\varphi_i=3.4$, which corresponds to $83$ $e$-folds of inflation. The trajectory is shown in the left panel of  Fig.~\ref{fig:traj_exp}. Our baseline parameters in Eq.~\eqref{eq:baseline_parameters} are chosen so that
 \begin{equation}
 	\ln\,10^{10}\,A_s=3.043\,,\,\,\,\,\,\,\, 	n_s=0.9618\,,\,\,\,\,\,\,\,\alpha_s=3.1\times10^{-6}\,,\,\,\,\,\,\,\, 
 	r_{0.002\,{\rm Mpc}^{-1}}=0.01\,,
 \end{equation}  well consistent with Planck/BICEP/Keck Array latest constraints at large scales. Note that the amplitude $\ln\,10^{10}\,A_s$, the tilt $n_s$  and the running of the tilt $\alpha_s$ are computed at the pivot scale $k_*=0.05\, {\rm Mpc}^{-1}$ assuming that it crosses the Hubble radius $N_*=55$ $e$-folds before the end of inflation\footnote{$N_*$ is affected by uncertainties that comes from our incomplete understanding of reheating. Clearly a larger/smaller $N_*$ would shift the peak in the power spectrum towards smaller/larger scales and give a bluer/redder $n_s$ compared to what we report in the main text.}. Changing some of the parameters in~ \eqref{eq:baseline_parameters} also slightly affects predictions at CMB scales. This effect can always be counterbalanced by tweaking more parameters at the same time. 
 
 We note that recent efforts to solve the so-called $H_0$ {\em tension} require a reinterpretation of available data, which, taken at face value, would imply higher values of $n_s$, all the way up to $n_s=1$, i.e. the
Harrison-Zeldovich value~\cite{Braglia:2020bym,Jiang:2022uyg,Cruz:2022oqk,Giare:2022rvg}. While it is not our goal to discuss such claims, we show at the end of this Section that we can easily obtain larger values of $n_s$ in our model.
 
 We will also consider another example described by the parameters: 
\begin{equation} 
	\label{eq:baseline2_parameters}
	M=9.48\times10^{-6}\,,\,\,\,\,\, 		\alpha=1\,,\,\,\,\,\, 		\tilde{g}=1\,,\,\,\,\,\, 		\tilde{m}=1\,,\,\,\,\,\, 		\chi_0=2.58\,,\,\,\,
	d=-10^{-5}\,,
\end{equation}
 Its trajectory is shown in the right panel of Fig.~\ref{fig:traj_exp}. As can be seen, compared to the baseline model, the hybrid field starts to move only when $\varphi$ has approached $\varphi\simeq0$. While this example has a spectral index which is too red to fit Planck data, the waterfall phase here is closer to a single-field stage of inflation, so it is useful for illustrative purposes and to be compared to the intuitive arguments presented in the previous Section.

Let us briefly summarize here the mechanism that leads to a peak in $\mathcal{P}(k)$,  also anticipated in the previous Sections,  from the perspective of the full multifield dynamics. The piece of the trajectory producing the perturbations at the CMB scales consists in a simple slow-roll phase along the $\varphi$ direction. During this stage, the  field $\chi$ slowly increases due to the linear term. Inflation is dominated by the slowly-rolling inflaton $\varphi$ and curvature perturbations do not experience amplification. Eventually, when $\varphi$ crosses the critical value $\varphi_c$, the effective mass squared  of the  field $\chi$ becomes negative, it develops an instability and isocurvature perturbations start to grow exponentially. 

Even during this period, however, curvature perturbations experience the usual evolution and freeze to a constant value after Hubble crossing. The trajectory turns and the  field $\chi$ rolls towards the minimum of the potential at $\chi_0$, taking a few $e$-folds to terminate inflation. At the moment of the turn, the coupling between curvature and isocurvature perturbations is switched  on and power in the latter is quickly transferred to the former, leading to a large peak in the PPS. We emphasize that the sourcing entirely occurs on super-Hubble scales, as opposed to recently proposed models where the turn sources amplification around Hubble crossing~\cite{Fumagalli:2020adf,Palma:2020ejf,Braglia:2020eai,Braglia:2020taf,Iacconi:2021ltm,Kallosh:2022vha}. It is also important to understand that, while isocurvature perturbations do grow during inflation, almost all of the power is transferred to the curvature ones during the turn, leaving the spectrum of primordial perturbations after inflation adiabatic, consistently with tight bounds on primordial isocurvature from CMB data~\cite{Planck:2018jri}.

To understand this process, it is useful to investigate the behavior of two important quantities, i.e. the effective mass squared of isocurvature modes on superhorizon scales:\footnote{We note that sometimes the quantity $m_{\rm iso}^2/H^2=m_{\rm eff,\,iso}^2/H^2-4\eta^2_\perp$ is also discussed in the literature. The two quantities are conceptually different as $m_{\rm eff,\,iso}^2/H^2$ is the {\em effective} mass squared of isocurvature perturbations obtained integrating out the evolution of curvature perturbations on super-horizon scales, while $m_{\rm iso}^2/H^2$ really corresponds to the mass of isocurvature perturbations. However, in hybrid inflation, most of the growth of isocurvature perturbations occurs {\em before} the turn, when $\eta_\perp^2\simeq0$ and the two quantities are numerically equivalent.}

\begin{equation}
	\label{eq:m_iso}
	\frac{m_{\rm eff,\,iso}^2}{H^2} = \frac{1}{\left(\dot{\varphi}^2+\dot{\chi}^2\right)\,H^2}\left[V_{\varphi\varphi}\,\dot{\chi}^2-2V_{\chi\varphi}\,\dot{\varphi}\dot{\chi}+V_{\chi\chi}\,\dot{\varphi}^2 \right]+3\eta^2_\perp,
\end{equation}
which describes the exponential growth of isocurvature modes, and the turn rate
\begin{equation}
	\label{eq:turn_rate}
	\eta_\perp^2=\frac{1}{\left(\dot{\varphi}^2+\dot{\chi}^2\right)\,H^2}\left[V_{\varphi}\,\dot{\chi}-V_{\chi}\,\dot{\varphi} \right]^2,
\end{equation}
which controls the coupling between curvature and isocurvature modes and the resulting amplification of the former. 
\begin{figure}[h!]
 \vskip -10pt
	\begin{center}
		\includegraphics[width=.495\columnwidth]{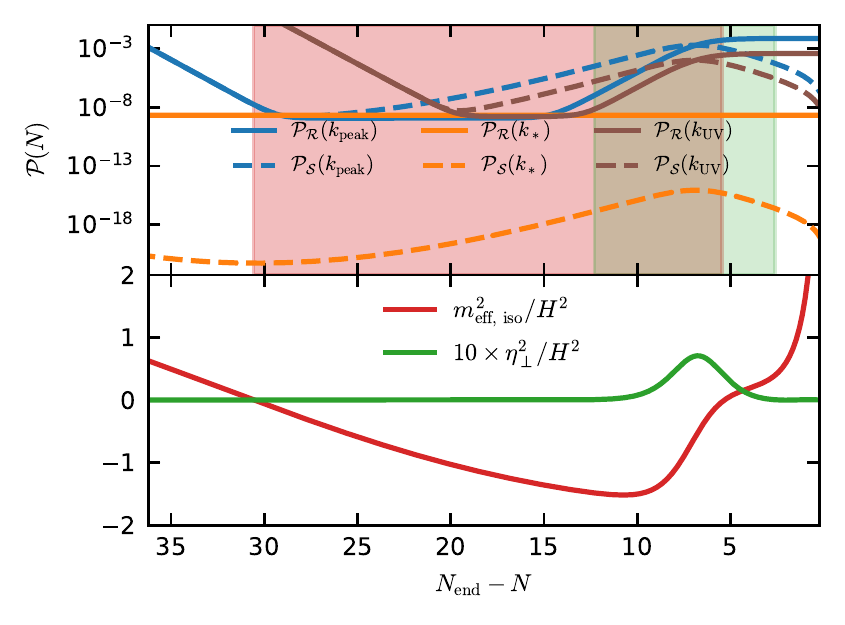}
		\includegraphics[width=.495\columnwidth]{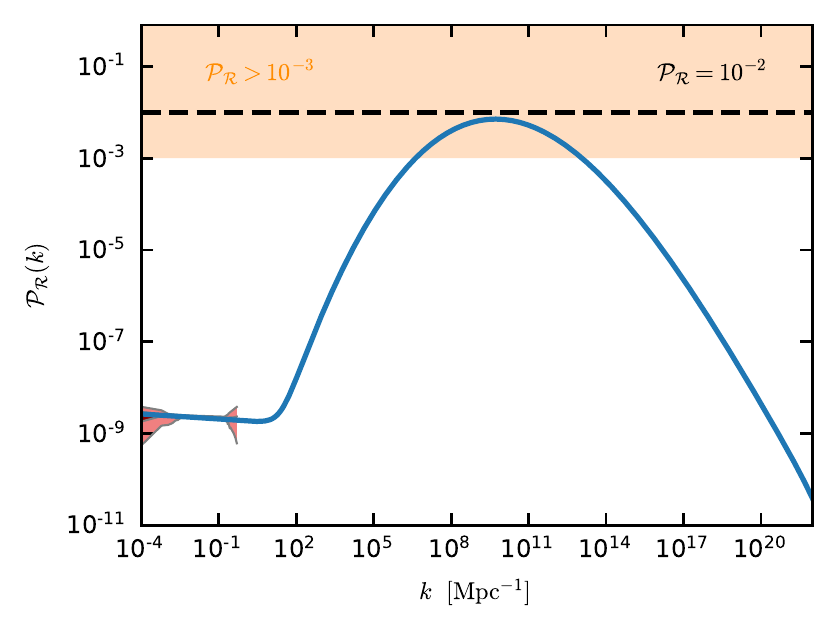}\end{center}
		 \vskip -15pt
	\caption{\footnotesize	\label{fig:baseline}  [Left] Time evolution of the dimensionless spectra of curvature and isocurvature perturbations in the baseline case. We show the evolution for the peak scale $k_{\rm peak}\simeq5.3\times 10^{9}\,{\rm Mpc}^{-1}$ and, for a comparison, the pivot scale $k_*=0.05\,{\rm Mpc}^{-1}$ and a smaller-scale mode $k_{\rm UV}= 10^{14}\,{\rm Mpc}^{-1}$. Green and red areas correspond to the regions where $\eta^2_\perp/H^2>10^{-2}$ and $m_{\rm eff,\,iso}^2/H^2<0$ respectively, as shown in the lower part of the plot.  [Right] PPS at the end of inflation. The yellow-shaded area is the region $\mathcal{P}_\mathcal{R}(k)\gtrsim 10^{-3}$ where we expect a sizable production of PBHs. The dashed black line is $\mathcal{P}_\mathcal{R}=0.01$, which is the value commonly used as criterium for PBH formation. The parameters used to produce this figure are the baseline parameters in~\eqref{eq:baseline_parameters}.}
\end{figure}

These quantities are plotted in the left panel of Fig.~\ref{fig:baseline}, together with the evolution of the dimensionless spectra of curvature $\mathcal{R}_k$  and isocurvature perturbations $\mathcal{S}_k$ for some illustrative wavenumbers $k$.
As can be seen,  as soon as $\varphi$ crosses $\varphi_c$ and $m_{\rm eff,\,iso}^2/H^2$ becomes negative,  isocurvature perturbations start to grow. This amplification is experienced by {\em all}  the modes, however, since  $m_{\rm eff,\,iso}^2/H^2>0$ for $\varphi>\varphi_c$, isocurvature modes which cross the Hubble radius long before the instability decay before have sufficiently decayed so that their amplitude is too small to source curvature perturbations, even if they get amplified (see dashed orange line in the left panel of Fig.~\ref{fig:baseline}). Note that isocurvature modes that cross the horizon later than the peak scale and are still sub-horizon at $\varphi=\varphi_c$) have less time to grow, resulting in a smaller amplification of curvature modes during the turn (see brown lines in Fig.~\ref{fig:baseline}). Very small scale modes cross the horizon long after the turn, when inflation is completely driven by $\chi$, and their evolution is just the typical slow-roll one, with isocurvature perturbations decaying straight after horizon crossing.

The net effect of such  amplification is to produce a broad bump in the PPS. If the amplitude of the bump is large enough, the perturbations can collapse into PBHs when they re-enter the Hubble radius during radiation dominated era~\cite{Garcia-Bellido:1996mdl}. The criterium for PBHs to form is very sensitive to many assumptions, and it is not the purpose of this paper to study it in details. Nevertheless, it generically requires $\mathcal{P}_\mathcal{R}\gtrsim10^{-3}-10^{-2}$~(see e.g. Refs.~\cite{Sasaki:2018dmp,Carr:2021bzv,Escriva:2022duf}) as indicated by the yellow shaded area in the right panel of Fig.~\ref{fig:baseline}.  This can easily be satisfied by our model. 

 The next issue to consider in the masses of PBH formed after inflation. This is a rather complicated story, we will limit ourselves to simple estimates. Consider perturbations produced at the peak, $\Delta N = N_{\rm end}-N_{\rm peak} $ e-foldings from the end of inflation. The mass of PBH can be estimated by the total energy inside  the horizon when these perturbations re-enter the horizon.  If the evolution until that time was matter dominated (which happens if reheating is very inefficient), then one can  roughly estimate  
$M_{\rm PBH} \sim O(10)\, H^{{-1}} e^{{3 \Delta N}}$, in Planck mass units, where $H$ is the Hubble constant at the end of inflation. For a radiation dominated universe (efficient reheating) the masses are expected to be much smaller, $
M_{\rm PBH} \sim O(10)\, H^{{-1}} e^{{2 \Delta N}}$~\cite{Garcia-Bellido:1996mdl,Linde:2012bt}.
The reason for the smaller mass in the second case is the redshift of radiation, which occurs prior to the horizon re-enter. In more realistic situations, where the PBH formation occurs in a universe with $p =\rho/3$ after a long stage of matter domination, the redshift is less efficient, and one may expect $M_{\rm PBH}$ between the  two limits discussed above \cite{Dalianis:2018frf,Iacconi:2021ltm}.  
\begin{figure} 
	\begin{center}
		\includegraphics[width=.495\columnwidth]{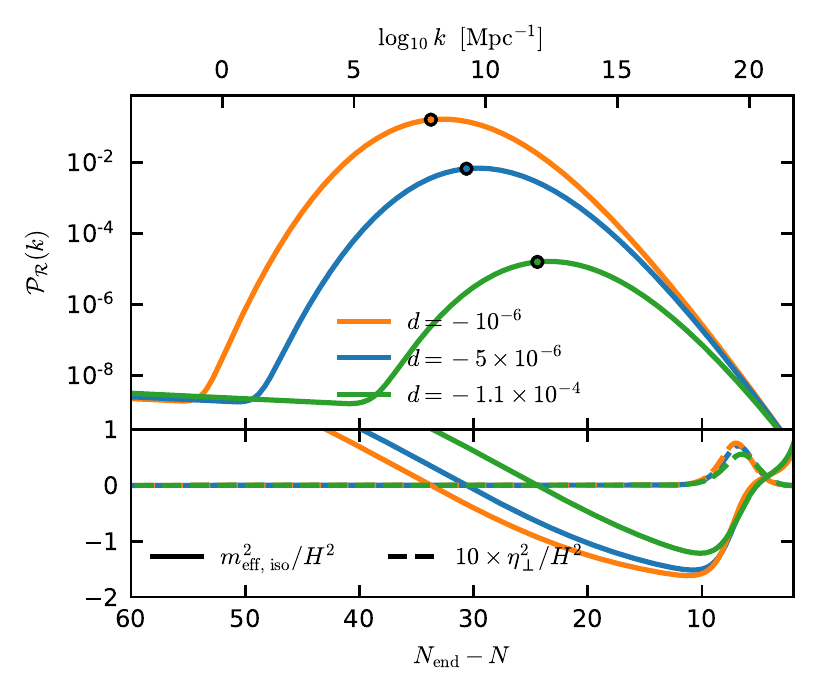}
		\includegraphics[width=.495\columnwidth]{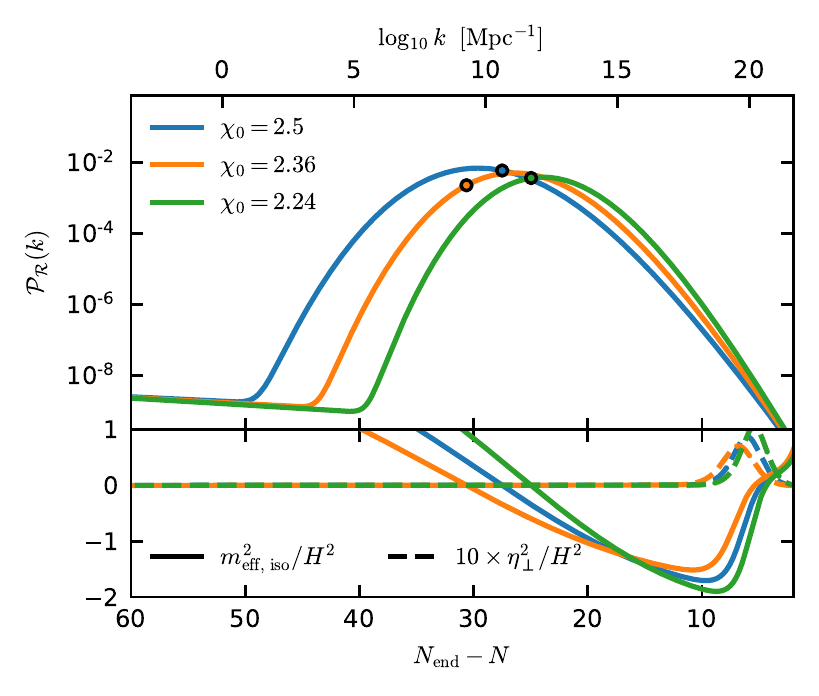}
		\includegraphics[width=.495\columnwidth]{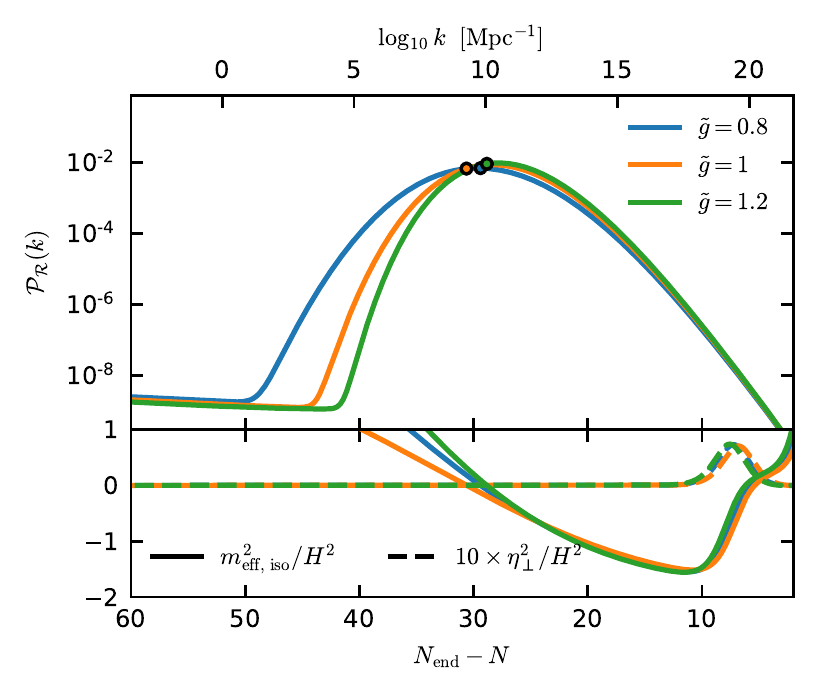}		\includegraphics[width=.495\columnwidth]{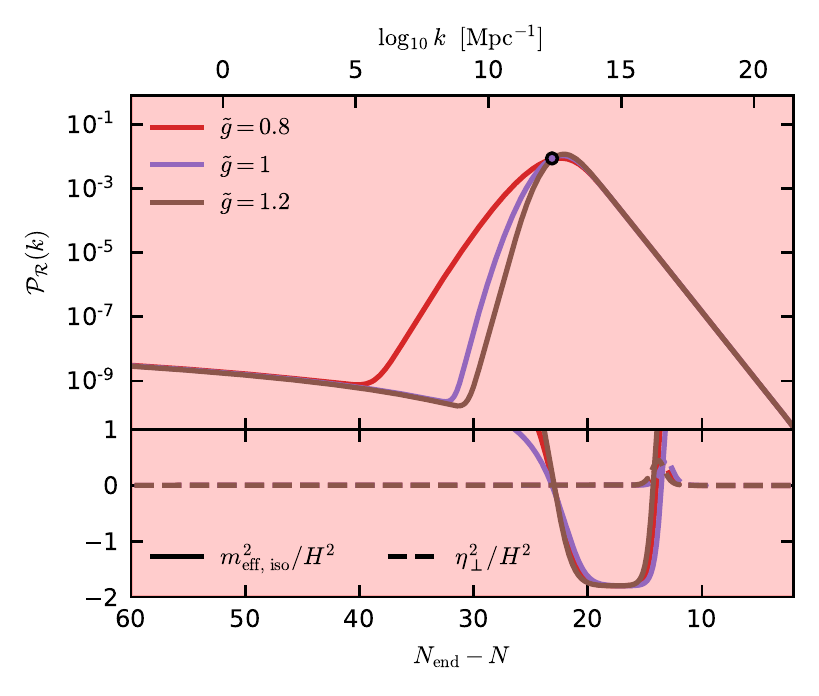}
	\end{center}
	\caption{\footnotesize	\label{fig:PK} Variation of model parameters and their imprints on the PPS. For each case, we also plot the evolution of $m_{\rm eff,\,iso}^2/H^2$ and $\eta^2_\perp/H^2$.  [Top-left] Variation of $d$, which controls the height of the peak.   [Top-right] Variation of $\chi_0$, which controls the location of the peak.   [Bottom-left] Variation of $\tilde{g}$, which controls the width of the peak.    [Bottom-right] Variation of $\tilde{g}$ for the parameters in~\eqref{eq:baseline2_parameters}. To highlight that the last case is not compatible with Planck observations, we plot it on top of a red background ($n_s$ is redder in that case). We also note that the x-scale axes on the bottom $N_{\rm end}-N$ correspond to the time at which the wavenumbers on the top axes cross the horizon i.e. $k=a(N_{\rm end}-N) H(N_{\rm end}-N)$.}  
\end{figure}

 To give a particular example, consider the universe which was radiation dominated soon  after the end of inflation. In this case one may 
use the expression that relates the scale of a perturbation to the mass of the formed PBH \cite{Wang:2019kaf}:
\begin{equation}
	\label{eq:k_to_m}
	\frac{M(k)}{M_\odot}\simeq10^{-16}\left(\frac{\gamma}{0.2}\right)\left(\frac{g(T_k)}{106.75}\right)^{-1/6}\left(\frac{k}{10^{14}\,{\rm Mpc}^{-1}}\right)^{-2},
\end{equation}
where $\gamma$ is a factor that encodes the efficiency of the collapse to PBHs and $g(T_k)$ is the effective number of relativistic degrees of freedom at the time (or, equivalently, temperature) of formation. For example, assuming $\gamma=0.2$ and $g(T_k)=106.75$, the baseline power spectrum would produce a  PBH mass function peaked around $M\sim10^{-8}\,M_\odot$. While this is excluded by microlensing experiments~\cite{Escriva:2022duf}, we will shortly see that the parameters can be tuned to shift the location of the peak and give rise to much  heavier or  much   lighter PBHs.  For example, the green line in the top-right panel of Fig.~\ref{fig:PK} shows a peak at $k_{\rm peak}=1.3\times10^{12}\,{\rm Mpc}^{-1}$, corresponding to PBHs of masses $M_{\rm PBH}\sim 6\times10^{-13}\,M_\odot$, that can constitute the totality of the Cold Dark Matter in our Universe~\cite{Escriva:2022duf}.   
 Also note that the abundance of PBHs depends exponentially on ratio of the primordial power spectrum to the critical value of the density contrast $\delta_c$, therefore, depending on the specific value adopted, the power spectrum shown in Fig.~\ref{fig:baseline} could result in no PBHs to an overproduction of PBHs.

The bump-like shape of the power spectrum at small scales suggests that we can control properties such as its amplitude, location and width by choosing different combinations of the model parameters. Phenomenologically, this does not strongly impact the formation of PBHs, which, being a critical process, is mainly sensitive to the peak amplitude~\cite{Cole:2022xqc}, so the relevant features are just its amplitude and location. However, the frequency profile has implications on the  SGWB (see next Section), which is sensitive to a wider portion of the bump.  Therefore, we now analyze the effects of varying the model parameters and show our results in Fig.~\ref{fig:PK}.

\begin{figure}
\vskip -10 pt
	\begin{center}
		\includegraphics[width=.45\columnwidth]{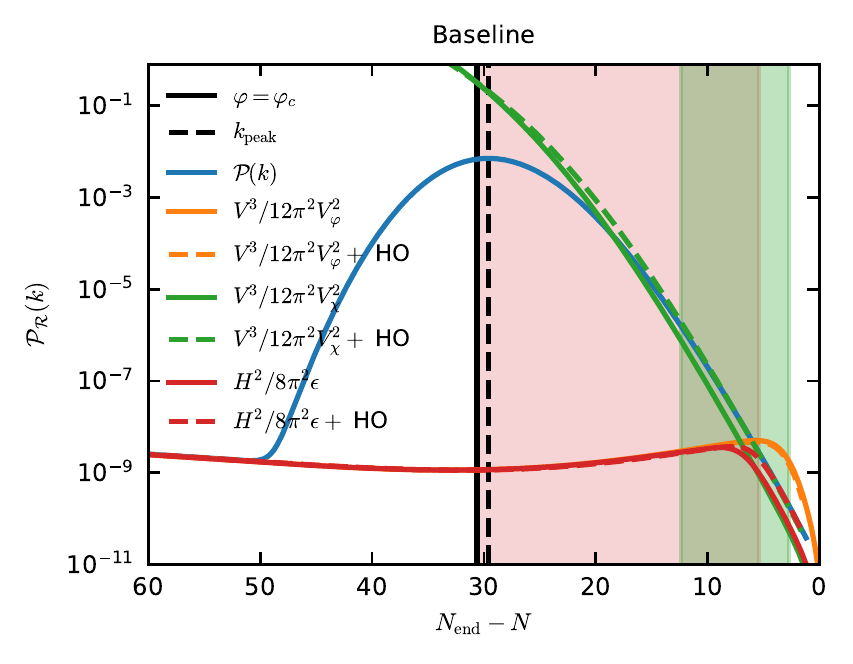}
		\includegraphics[width=.45\columnwidth]{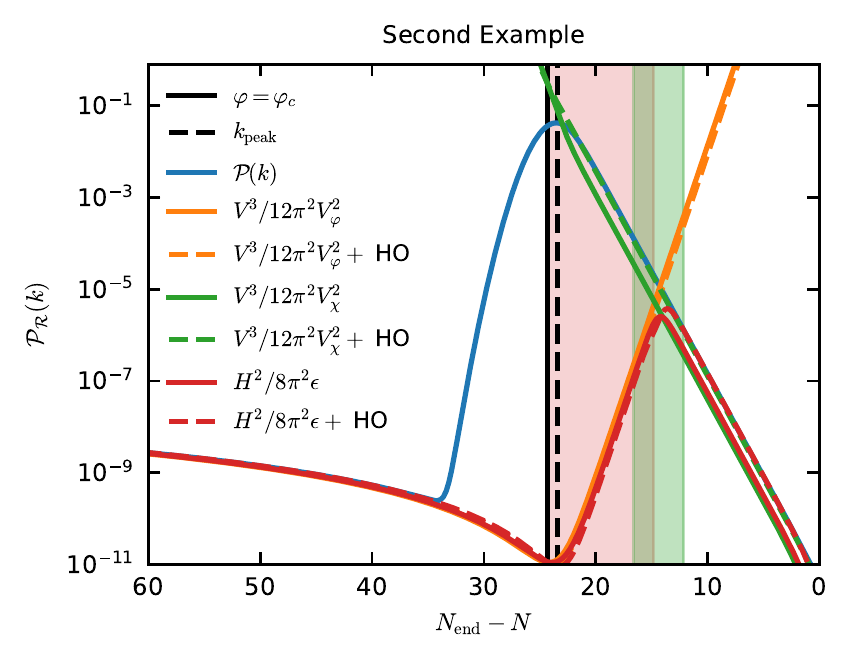}
	\end{center}
		\vskip -10 pt
	\caption{\footnotesize	\label{fig:1102}  We compare the single-field slow-roll approximations for the primordial power spectrum to the exact numerical results for the baseline example in Eq.~\eqref{eq:baseline_parameters} (left) and for the example in Eq.~\eqref{eq:baseline2_parameters} (right). Red and green regions have the same meaning as in Fig.~\ref{fig:baseline}. }
	\label{fig:comparison}
\end{figure}

The parameter controlling the {\em amplitude of the peak} is the amplitude $d$ of the linear term in the potential. Intuitively, this is very simple to understand. As we have shown in the previous section, in the absence of the linear term in the models with $\chi_{0} \gtrsim 1.7$ one has  eternal inflation at the ridge of the potential at $\chi = 0$, which results in $\mathcal{O}(1)$ perturbations. The term $d\,\chi$ moves us away from the eternal inflation regime, decreasing the amplitude in a controllable way. When the field $\vp$ moves sufficiently far away of the critical point $\vp_{c}$, its equations of motion depend less and less on the field $\vp$.  Following the argument in the previous Section, we would like to approximate the resulting amplitude of perturbations produced at this stage  by the simple single-field equation \rf{as}. We illustrate it in Fig. \ref{fig:comparison}.   

As can be seen, the approximation is quite bad for the amplitude of the peak in the baseline example though it becomes reasonable for the spectrum away from the peak.  This approximation  becomes much better  in the second example, shown in the right sides of Figures~\ref{fig:traj_exp}, \ref{fig:comparison}. The reason is that in the second example the turn is sharper and the waterfall stage is entirely driven by $\chi$, i.e. the trajectory is in the $\chi$ direction, with the field $\varphi$ being approximately constant $\varphi\simeq0$. Therefore, during the last $\sim10$ $e$-folds after the turn  with $\eta_\perp^2/H^2\sim0$, the numerical power spectrum matches well the slow-roll prediction $\mathcal{P}_\mathcal{R}\simeq V^3/12\pi^2 V_\chi^{2}\simeq H^2/8\pi^2\epsilon$. On the other hand, in the baseline case the trajectory is still slowly turning close to the end of inflation, and the approximation to the single-field inflation driven by $\chi$ is rough.

We note that, though $\epsilon$ only becomes large close to the end of inflation, it does grow reasonably fast during the waterfall stage, so its logarithmic derivative, i.e. the second slow-roll parameter $\eta$ is not negligible. Therefore the simple expressions $\mathcal{P}_\mathcal{R}\simeq V^3/12\pi^2 V_{\chi\,(\varphi)}^{2}\simeq H^2/8\pi^2\epsilon$, which are derived at first order in slow-roll receive higher order corrections. In particular, the expression for the spectrum at horizon crossing $k\sim a H$ becomes~\cite{Schwarz:2001vv}:
	
\begin{equation}
\mathcal{P}_\mathcal{R}\simeq\frac{H^2}{8\pi^2\epsilon}\left[\frac{1}{\pi}\left(2-2\epsilon\right)^{2\nu-1}\,\Gamma^2(\nu)\right],
\end{equation}	
where	$\nu= 1/2 \,+\,1/(1-\epsilon)\,+\,\eta/2$ and $\Gamma$ is the Gamma function. Expressions for this quantity during the stages of inflation driven by $\varphi$ or $\chi$ can be obtained by using $\epsilon\simeq\epsilon_V=\left(V_{\varphi\,(\chi)}/V\right)^2\,/\,2$ and $\eta\simeq4\epsilon_V-2 \eta_V$, where $\eta_V= V_{\varphi\varphi\,(\chi\chi)}/V$, where $\phi_i$ denotes either $\chi$, or $\phi$, depending on which one drives inflation at a given stage.

As can be seen from the dashed lines in Fig.~\ref{fig:1102}, the agreement on the spectrum produced after the peak is now much better and the formula correctly reproduces the power spectrum at the very small scales that cross the horizon when the slow-roll parameters are large and the hybrid field $\chi$ is ending inflation.

 Another interesting feature that can be understood from Fig.~\ref{fig:comparison} is that the peak in the power spectrum is very close to the scale that crosses the horizon when $\vp=\vp_c$, i.e. the one for which isocurvature perturbations experience the largest growth  without having decayed on superhorizon scales prior to  $\vp=\vp_c$. While, in the second example, we see from the plot that the amplitude of the spectrum at $k_{\rm peak}$ can be approximated quite well by $\mathcal{P}_\mathcal{R}\simeq V^3/12\pi^2 V_\chi^{2}$, we stress again, that this approximation is rigorous only after the turn, when inflation is effectively single-field, and that we must compute the full multifield dynamics to get exact results.

The {\em location of the peak} is controlled by the duration of the waterfall phase, which essentially depends on the $\chi_0$ (see top-right panel of Fig.~\ref{fig:PK}). Smaller values of $\chi_0$ imply that the minimum in the $\chi$ direction is closer to $\chi\simeq0$ and the waterfall stage is thus shorter. For this reason, the amplified modes cross the Hubble radius closer to the end of inflation and the peak in the power spectrum occurs at smaller scales (larger wavenumbers).  Therefore, tuning $\chi_0$, we can easily control  the mass of the produced PBHs, which  scales exponentially with $N$, i.e. $M\sim k^{-2}\sim e^{2 (N_{\rm end}-N_{\rm peak})}$, see Eq.~\eqref{eq:k_to_m}. Clearly, shifting $\chi_0$, we would need to adjust other parameters to maintain the consistency with large scale data, as explained below.

Finally, the coupling between the two fields $\chi$ and $\varphi$ is responsible for the {\em width of the peak}. While the coupling does not change the stage with $m_{\rm eff,\,iso}^2/H^2<0$, it significantly affects the evolution of $m_{\rm eff,\,iso}^2/H^2$ at earlier times. In particular, a larger value of $g$ makes the effective squared mass of isocurvature perturbations much larger for $\varphi>\varphi_c$. Therefore, isocurvature modes that are super-horizon before the tachyonic instability decay very fast and their subsequent amplification is not enough to impact the evolution of curvature perturbations. On the other hand, modes that cross the Hubble radius after the tachyonic instability has developed are not affected by this. As such, the shape of the bump is only modified at scales $k<k_{\rm peak}$  (see bottom-left panel of Fig.~\ref{fig:PK}). This effect is particularly pronounced in the bottom-right panel, where we show the variation with $\tilde{g}$ for a different combination of parameters.

 Armed with our understanding of the model parameters and their effects on the power spectrum, we would like to ask a final question. What is the relation of our results to the simple $\alpha$-attractor prediction\footnote{We only consider the spectral index here for simplicity.}  
$n_s=1-\frac{2}{N}$?  Or, in other words, how much are we taking advantage of the attractor nature of the model?

 As we will shortly discuss, these questions are partly related to the last parameter of the model, whose effects we have not explored so far, i.e. the parameter $\tilde{m}$, which controls the amplitude of the plateau at large values of $\varphi$ relative to the uplift. We note that the following discussion can be straightforwardly generalized to {\em polynomial} attractors model that we construct in Section~\ref{sec:polynomial}, which have different attractor predictions, see e.g. Eq.~\eqref{genns}.

Let us first clarify the meaning of $N$ in $n_s(N)=1-\frac{2}{N}$. In single-field $\alpha$ attractors models, $N$ represents the number of $e$-folds from the moment at which the pivot scale crosses the Hubble radius to the end of inflation. Therefore, to be as clear as possible, the quantity we are interested is $n_s(N_*)$. As stressed above, we adopt $N_*=55$ in our paper. More specifically, the attractor expression is derived by analytically solving the slow-roll equation for the inflaton $\varphi(N)$ as a function of $N$ and plugging it into the slow-roll approximation for the spectral index $n_s\simeq1-6\epsilon_V+2\eta_V$. 

However, in our model, where we have two stages of inflation, the equation for $n_s$ needs to be modified as~\cite{Iacconi:2021ltm,Kallosh:2022ggf}: 
\begin{equation}
	\label{eq:ns_attr_modified}
	n_s=1-\frac{2}{N_*-\Delta N}.
\end{equation} 
This is very easy to understand, as the second stage shifts the end of inflation by $\Delta N$. In this way, we relate the value of $n_s$ to the region of the plateau that is producing the perturbations at $N_*$. However, how do we define $\Delta N$? An intuitive way to do so would be to define it as the distance from the time where $\eta_\perp^2/H^2$ peaks, which roughly separates the region two stages of inflation driven $\varphi$ and $\chi$, to the end of inflation. However, in our model,  there are situations where this is not the best definition for $\Delta N$. Take the second example in Fig.~\ref{fig:traj_exp}. There, $\varphi$ approaches $0$ some $e$-folds after $\epsilon_\varphi= (V'/V)^2 /2$ reaches its maximum, i.e. the point of maximum velocity of $\varphi$,  and significantly slows down afterwards. However, since $\chi$ is moving very slowly and its kinetic energy is much smaller than that of $\varphi$, it takes roughly $\sim 10$ $e$-folds before $\eta_\perp^2/H^2$ peaks and the trajectory turns into the $\chi$ direction. This is  different from what would happen in the single-field case, where inflation would end shortly after $\epsilon_\varphi$ reaches its maximum value, which in that context is $\epsilon_\varphi\,=\,1$.  

A more sensible definition, which resembles more the one single-field $\alpha$-attractor model, is in fact to choose the time at which $\epsilon_\varphi$ peaks, so that $\Delta N \equiv N_{\rm \epsilon_\varphi\,=\,{\rm maximum}} - N_{\rm end}$\footnote{We stress that this is $\Delta N$ is not the duration of the waterfall stage, which instead begins when $\varphi=\varphi_c$ and $\chi$ develops an instability. }. In this way, regardless of what happens after  $N_{\rm \epsilon_\varphi\,=\,{\rm maximum}}$, Eq.~\eqref{eq:ns_attr_modified} should match quite well the prediction, as long as the first stage of inflation is independent from the first one. This, in turn depends mainly on the combination of parameters $\tilde{m}\,\sqrt{\alpha}$.

Also, note that, even after the modification $N_*\to N_*-\Delta N$ the simple Eq.~\eqref{eq:ns_attr_modified} would be adequate only if there was no uplifting potential.  In that case, the equation needs to be further modified as:
\begin{equation}
	\label{eq:ns_final}
	n_s-1=1- \frac{2 V_0 \gamma^2}{V_0\gamma^2\left(N_*-\Delta N\right)+\left(V_{\rm up}+V_0\right)e^{\gamma s_{c}}}
,\end{equation}
where $\gamma=\sqrt{2/3\alpha}$, $s_{c}=\varphi_{c} -\frac{1}{\gamma}\ln 4$, $V_0= 3 \alpha\tilde{m}^2M^2$ and $V_{\rm up}=M^2\chi_0^2/4$,  see Eq.~(4.10) of Ref.~\cite{Kallosh:2022ggf}. Note that Eq.~\eqref{eq:ns_final}  reduces to Eq.~\eqref{eq:ns_attr_modified}   if   $  V_{\rm up}+V_0 \ll V_0\gamma^2 (N_*-\Delta N) e^{-\gamma s_{0}}  $. On the other hand,  in the large uplift limit   $  V_{\rm up}+V_0 \gg  V_0\gamma^2 (N_*-\Delta N)e^{-\gamma s_{0}}$ the value of  $n_{s}$ moves towards $n_{s }= 1$~\cite{Kallosh:2022ggf}.\footnote{Note that with an increase of $ V_{\rm up}$ the attractor $n_{s} = 1$ may be reached twice. Indeed, when $ V_{\rm up}$ increases, the evolution of the field $\vp$ slows down, so the last $55$ e-foldings of inflation may happen at $\vp < \sqrt{6\alpha}$, when the potential  $\sim \tanh^{2}{\varphi\over\sqrt {6 \alpha}} $  is reduced to the original quadratic potential. In this case, with the growth of $ V_{\rm up}$ the value of $n_{s}$ crosses $n_{s} = 1$, continues to grow for a while, and then decreases and approaches its final attractor value $n_{s} = 1$. And with the  account taken of the full  two-field evolution, the situation becomes even more complicated, which is why we needed a detailed numerical analysis described in this section.}

\begin{figure}
	\begin{center}
		\includegraphics[width=.47\columnwidth]{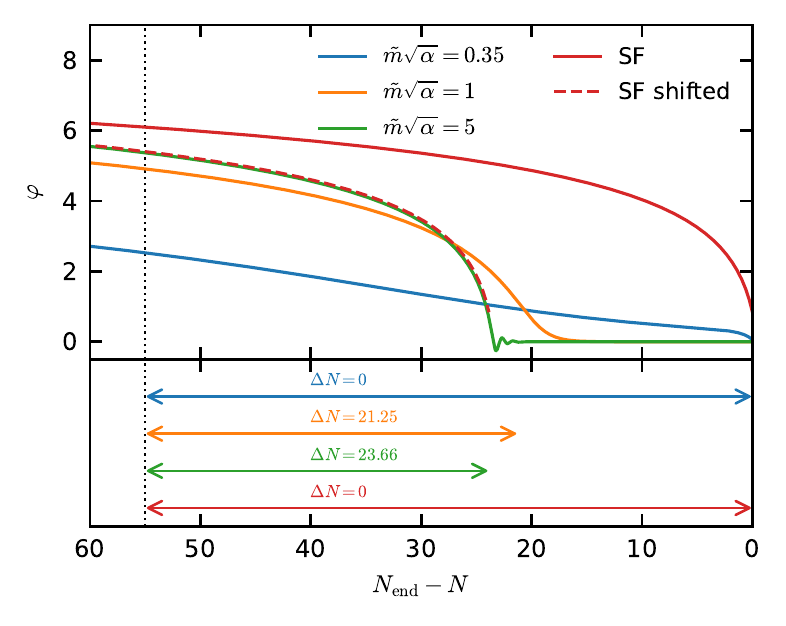}
		\includegraphics[width=.51\columnwidth]{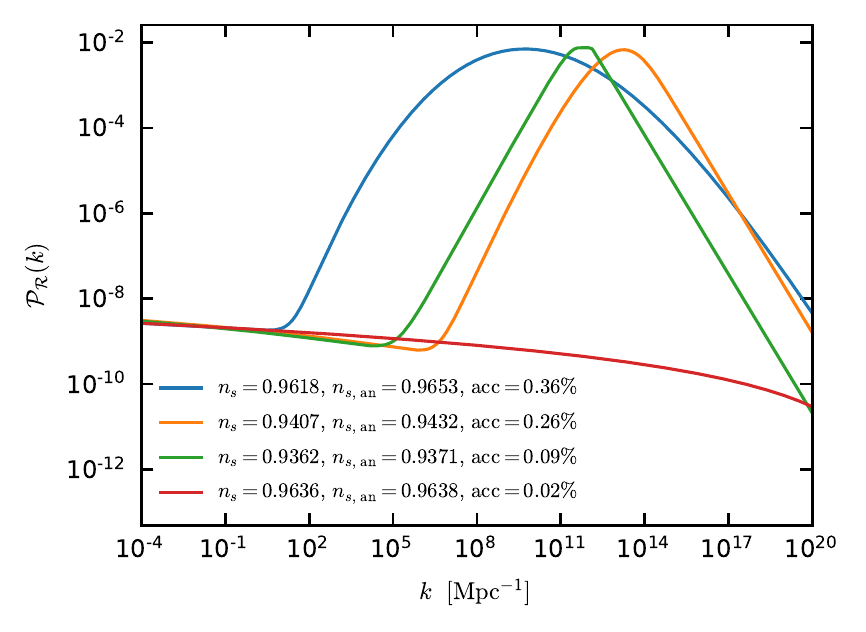}
	\end{center}
	\caption{\footnotesize	\label{fig:ns_m} Varying $\tilde{m}$ and its effect on the field evolution (left) and the primordial power spectrum (right).  }
\end{figure}
 
In Fig.~\ref{fig:ns_m}, we show the sensitivity of our results to the parameter $\tilde{m}$. We plot the evolution of the field $\varphi$ together with the corresponding primordial power spectra. The change in  $\tilde{m}\sqrt{\alpha}$ is accompanied by a change in the initial condition $\varphi_i$ to get the same number of $e$-folds in all the cases and by a change in the $d=-10^{-5}$ in the case with $\tilde{m}\sqrt{\alpha}=1$ and $\tilde{g}=2.5$ in the case with $\tilde{m}\sqrt{\alpha}=5$, so that all the spectra show a peak of a comparable amplitude. $\alpha$ and $\chi_0$ are kept fixed to the baseline values. For a comparison, we show in red the evolution of $\varphi$ in the single-field $\alpha$-attractor model.  

 As can be seen, the evolution for $\varphi$ gets more and more similar to the corresponding single-field evolution as $\tilde{m}\sqrt{\alpha}$ increases,  recovering the attractor predictions for large values of $V_0/V_{\rm up}$. This is confirmed by the values of $n_s$ reported in Fig.~\ref{fig:ns_m}, which agree with Eq.~\eqref{eq:ns_final}.  
 Therefore, the conclusion is that, if we want to take advantage of the attractor predictions, we need a large $V_0/V_{\rm up}$. In that case, however, for typical values of $\Delta N$ required to have interesting phenomenological effects at small scales, $n_s\sim 1-2/(N_*-\Delta N)$ is too red to fit Planck data.
	
	Let us stress that this does not mean that our model is excluded by observations. In fact, as discussed above, the baseline example perfectly agrees with Planck measurements. However, in that example, we must go slightly off   the attractor regime,  and compensate for the decrease of $n_{s}$ by an increase of $n_{s}$ due to the uplift by $V_{\rm up}=M^{2} \chi_{0}^{2}/4$~\cite{Kallosh:2022ggf}). 
	
	Things are different for  polynomial hybrid attractors  that we introduce in the next Section, where one does not need to rely on uplifting   for consistency with the Planck data. But for the exponential attractors discussed in this Section, the proximity of the attractor regime is also very helpful. For example, if $\Delta N = 10$ and $N_* \sim 55$, the attractor prediction ignoring the uplifting would be $n_{s} \sim 0.956$. This would inform us that with an account taken of uplifting the value of $n_{s}$ does not go  below $0.956$, and we only need an uplift  increasing  $n_{s}$ by less than $0.01$ to make it   consistent with the Planck results. It is always possible to do it; in the large uplift limit one can bring $n_{s}$ all the way up to $n_{s}  = 1$~\cite{Kallosh:2022ggf}.  This healing power of uplift may play a significant role in the development of advanced inflationary models compatible with the changing observational landscape, where the attempts to resolve the $H_{0}$ problem may push us towards considering inflationary models with large $n_{s}$ \cite{Braglia:2020bym,Jiang:2022uyg,Cruz:2022oqk,Giare:2022rvg}.

\begin{figure} 
	\begin{center}
		\includegraphics[width=.495\columnwidth]{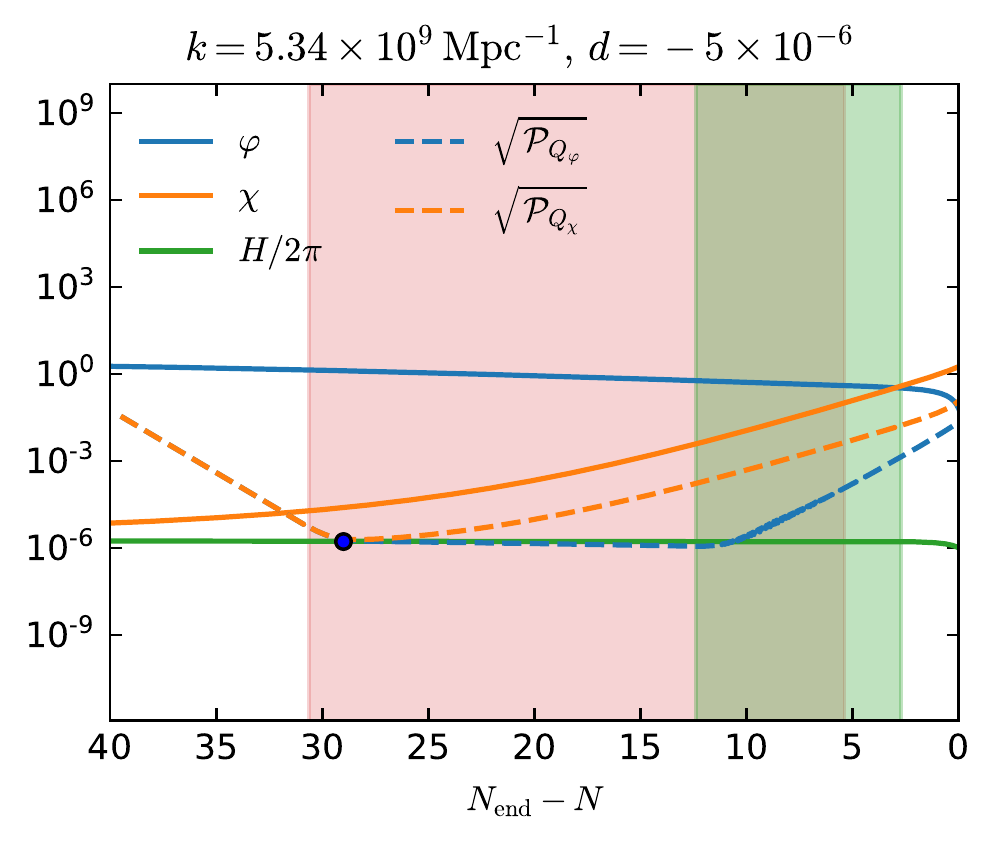}
		\includegraphics[width=.495\columnwidth]{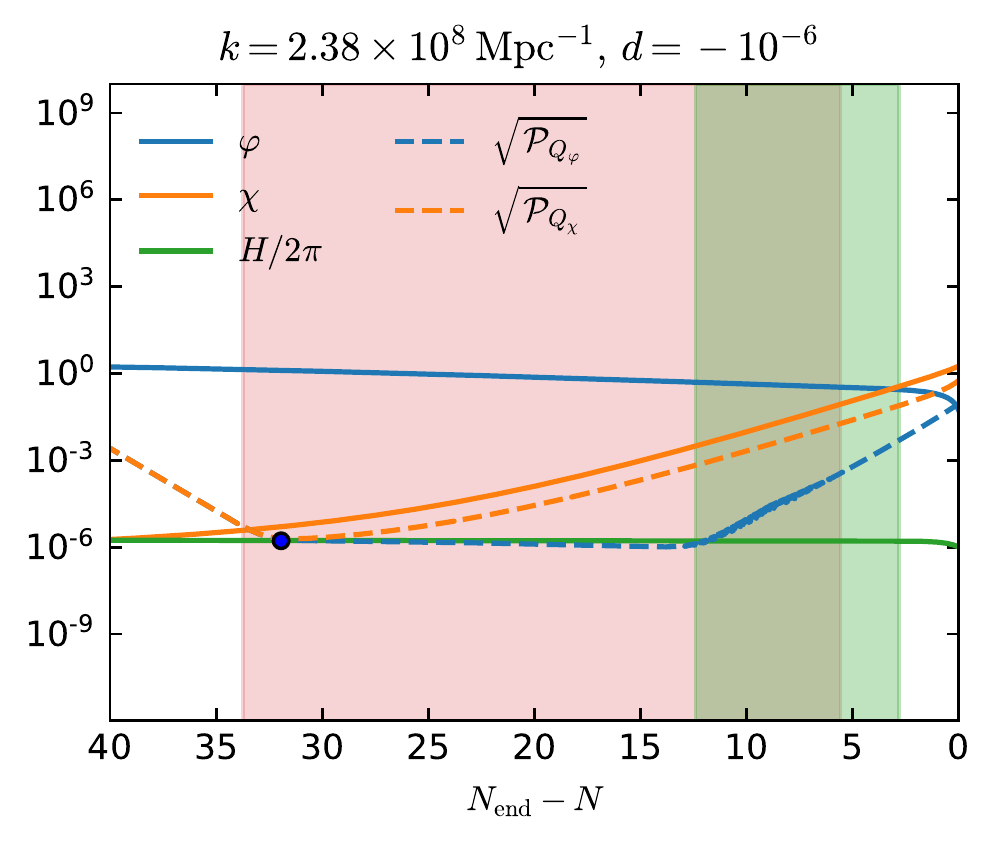}
	\end{center}
	\caption{\footnotesize	\label{fig:PertEvo} Evolution of perturbations for the cases shown in the top left panel of Fig.~\ref{fig:PK} with $d=-5\times10^{-6}$ (left) and $d=-10^{-6}$ (right). We plot the evolution for the mode corresponding to the peak in the power spectrum and we denote the horizon crossing with a small blue circle. The red and green shaded regions have the same meaning as in Fig.~\ref{fig:baseline}. }
\end{figure}

{\bf Growth of perturbations.} Before concluding, we note that since the perturbations in our model grow very large, care should be taken for them not to backreact on the background evolution. As a naive criterium, which should anyway give a good estimate of when backreaction is negligible, we compare the background values of $\varphi$ and $\chi$ to the evolution of the corresponding perturbations $\sim\sqrt{\mathcal{P}_{Q_\varphi}}$ and  $\sim\sqrt{\mathcal{P}_{Q_\chi}}$. Here, we denote the gauge-invariant perturbations to the two fields by $Q_\varphi$ and $Q_\chi$.
We show this in Fig.~\ref{fig:PertEvo}, where we plot the evolution of the $k_{\rm peak}$ modes, which is the one that takes larger values.

 In the left panel, corresponding to the baseline case, where $\mathcal{P}_\mathcal{R}(k_{\rm peak})\sim0.007$, we see that the perturbations are always subdominant with respect to the background fields. In particular, after the perturbation to $\chi$ crosses the horizon, where it takes the typical value $H/2\pi$, it starts growing because of the tachyonic instability, but we see that the background value for  $\chi$ is  always at least $\sim15$ times larger. We therefore conclude that in this case the perturbations should not affect the background evolution and our results are robust. As another example, we show the evolution for $k_{\rm peak}$ in the case of a smaller $d=-10^{-6}$, also shown in Fig.~\ref{fig:PK}. We see that also in this case perturbations remain small compared to the background value, although now $\sqrt{\mathcal{P}_{Q_\chi}}$ is only a factor of $\sim3$ smaller, so the two quantities are of the same order of magnitude. However, this case has a very large peak amplitude  $\mathcal{P}_\mathcal{R}(k_{\rm peak})\sim0.17$ so that it is excluded by the over-production of PBHs and also by current Pulsar Timing Array (PTA) limits on the SGWB induced by the scalar perturbations at second order (see next Section).

\begin{figure}
	\begin{center}
		\includegraphics[width=.52\columnwidth]{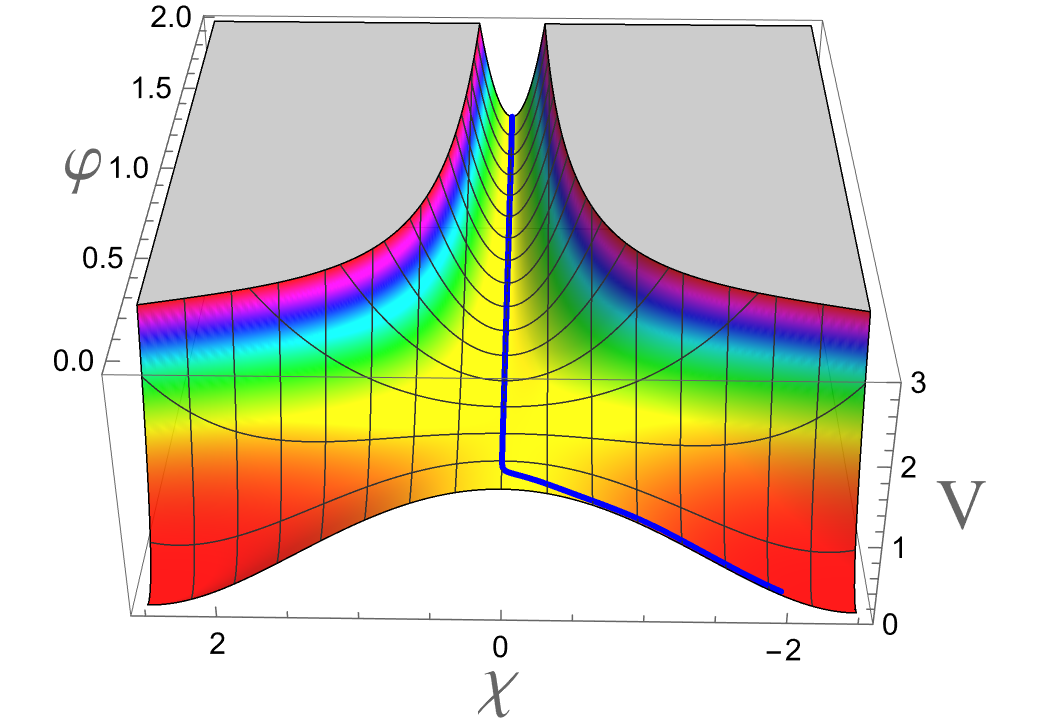}
		\includegraphics[width=.47\columnwidth]{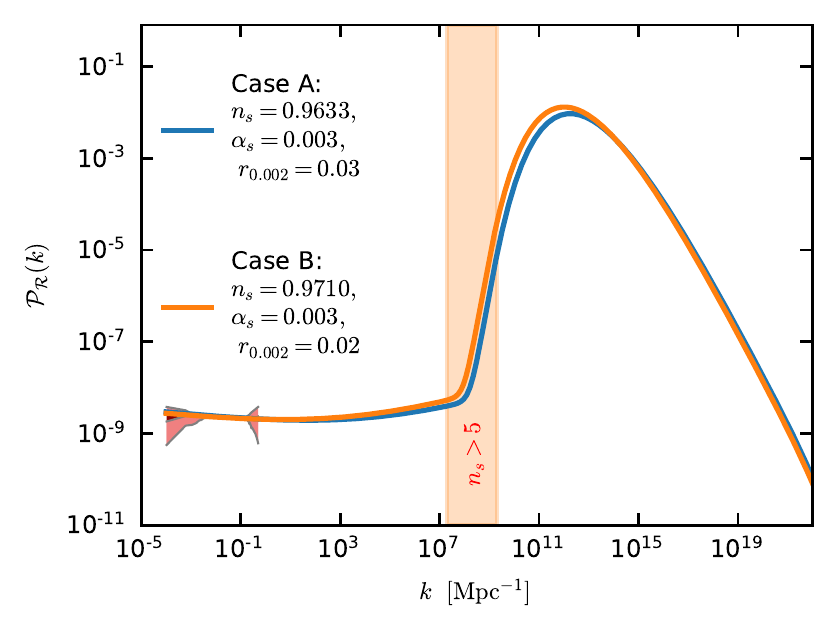}
	\end{center}
	\caption{\footnotesize	\label{fig:high_ns_exp}  Examples with a larger $n_s$. We use the following parameters. Case A:  $\alpha=0.9\,,\, 		 		\tilde{m}=0.35\,,\, 	\chi_0=2.5\,,\,\,\,\tilde{ g}=3.7\,,\,\,\,
		d=5\times10^{-6}\,$. Case B:  $\alpha=0.6\,,\, 		 		\tilde{m}=0.35\,,\, 	\chi_0=2.5\,,\,\,\, \tilde{g}=5\,,\,\,\,
		d=5\times10^{-6}\,$. The parameter $M$ is fixed to match the COBE normalization and $\varphi_i$ to achieve $~70$ $e$-folds of inflation. The potential is shown in units $M^{2}$. The trajectory in the left panel is shown for Case A, but is representative also of Case B, which shares the same features. }
\end{figure}

\paragraph{A larger $n_s$.} Although the example above is perfectly consistent with CMB observations, it is interesting to explore whether larger values of $n_s$ can be attained in our model. We show two examples in Fig.~\ref{fig:high_ns_exp}, where  $n_s$ falls both on the right of the sweet spot of the Planck/BICEP/Keck Array. Note that also a non-negligible amount of running is produced, which is nevertheless consistent with the constraints from Planck~\cite{Planck:2018jri}. As seen in the left panel, the main difference with respect to the baseline evolution is that, since $g$ is larger and $\varphi_c$ smaller, the hybrid field $\chi$ starts to move later during inflation, with a resulting shorter waterfall stage. The CMB scales therefore test the region where the plateau is flatter and $n_s$ bluer.  However, also in this case, the uplifting potential is quite large, i.e. the ratio $V_0/V_{\rm up}$ is small, and we cannot rely on the simple attractor relation $n_s=1-2/(N_*-\Delta N)$, as discussed above. 

We also note another interesting feature of the power spectrum in Fig.~\ref{fig:high_ns_exp}. The power spectra in the yellow-shaded region rises with a spectral index of roughly $10.9$ and $11.2$ for Case A and B respectively. This growth is much steeper than $n_s\sim5$ (plus a weak running) which is the limit in the case of canonical single field inflation~\cite{Byrnes:2018txb,Carrilho:2019oqg}, confirming that multiple field effects can easily overcome such limit \cite{Palma:2020ejf,Fumagalli:2020adf,Braglia:2020taf}.

\section{Stochastic Gravitational Wave background generated at horizon re-entry } 
\label{sec:SGWB}

It is well known that scalar and tensor perturbations are decoupled at linear order. However, they are mixed at higher orders in perturbation theory. In particular, scalar perturbations act as a source of gravitational waves at second order in perturbation theory when they re-enter the Hubble radius during the radiation dominated era\footnote{In principle, since the perturbations produced during inflation are quite large, in addition to the gravitational waves produced during the radiation era, we could also expect a sizable background to be sourced already during inflation~\cite{Biagetti:2013kwa}. However, as shown in Ref.~\cite{Fumagalli:2021mpc}, its amplitude is suppressed, relative to that produced after inflation, by powers of $\epsilon$  and $\eta_\perp/H$. Since both factors are much smaller than $1$ in our model, we ignore this contribution.}~\cite{Carbone:2004iv,Ananda:2006af,Baumann:2007zm}. If their amplitude is large enough, the resulting SGWB can fall into the sensitivity of future gravitational wave interferometers, that can therefore be used to test the physics of Hybrid inflation. It is therefore crucial to make accurate predictions on the spectral shape of the SGWB.

\begin{figure}
	\begin{center}
		\includegraphics[width=.495\columnwidth]{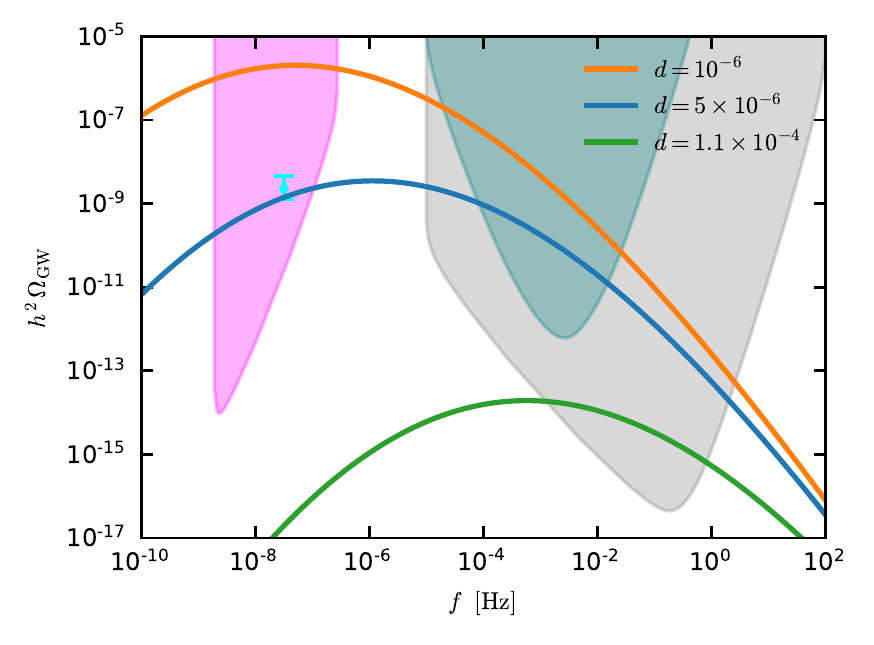}
		\includegraphics[width=.495\columnwidth]{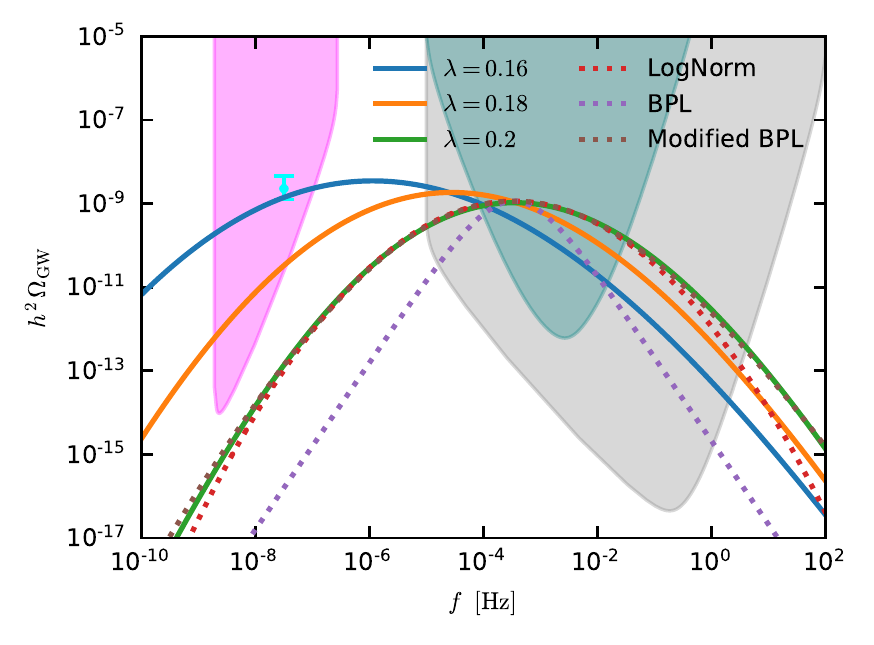}
		\includegraphics[width=.495\columnwidth]{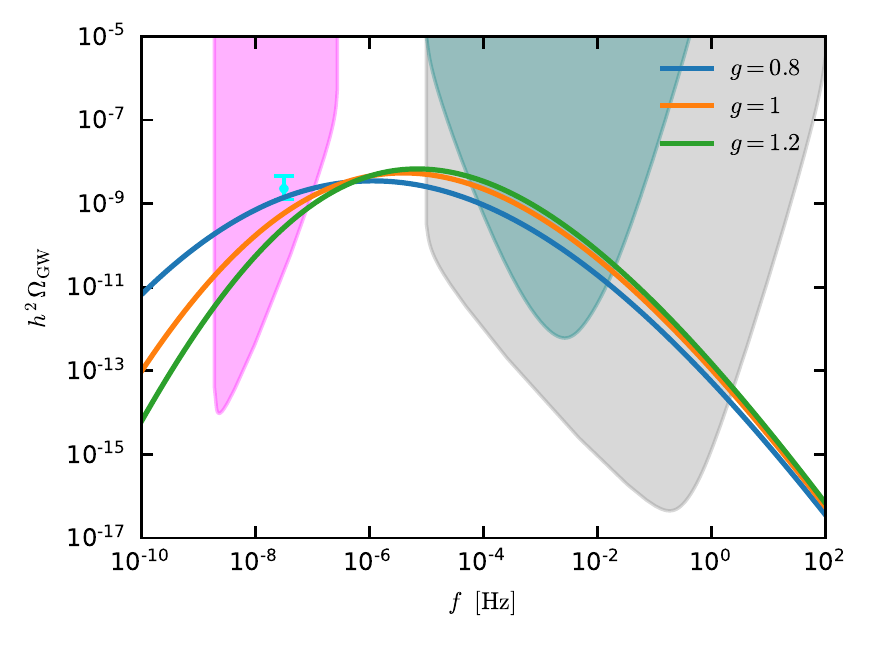}		\includegraphics[width=.495\columnwidth]{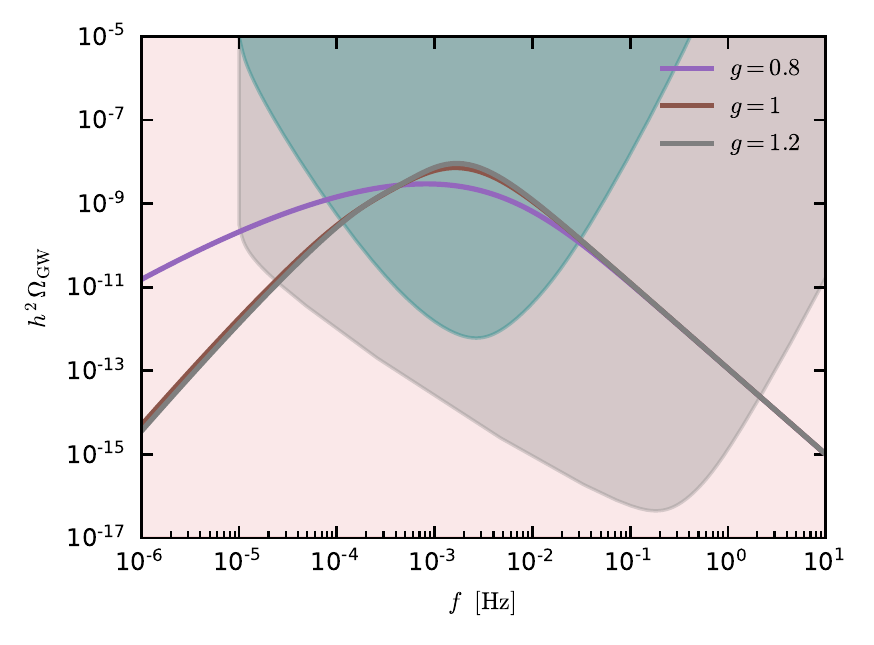}
	\end{center}
	\caption{\footnotesize	\label{fig:sgwb} SGWB sourced by the large amplitude perturbations. We consider the same parameters as in Fig.~\ref{fig:PK}.  Magenta, turquoise-green and grey shaded regions correspond to the sensitivities of SKA~\cite{Zhao:2013bba}, LISA~\cite{LISACosmologyWorkingGroup:2022jok} and BBO/DECIGO~\cite{Yagi:2011wg}.} 
\end{figure}

The energy density of the SGWB at present time is given by~\cite{Kohri:2018awv,Espinosa:2018eve}:
\begin{equation}
	\label{eq:sgwb}
	\Omega_{\rm ind}(k) = 0.387\, \Omega_{\rm R} \left(\frac{g_{*,s}^{4}g_{*}^{-3}}{106.75}\right)^{-\frac{1}{3}}
	\frac{1}{6} \int_{-1}^1 \dd x \int_1^\infty \dd y \, \mathcal{P}_\mathcal{R}\left(\frac{y-x}{2}k\right) \mathcal{P}_\mathcal{R}\left(\frac{x+y}{2}k\right) F(x,y) \,,
\end{equation}
where $\Omega_{\rm R} = 5.38\times10^{-5}$ is the radiation abundance~\cite{Planck:2018vyg}, the effective numbers of degrees of freedom, $g_{*,s}$ and $g_*$, are evaluated at the moment when the constant abundance is reached, roughly coinciding with the horizon crossing time, and
\begin{align}
	F(x,y) = &\frac{288(x^2+y^2-6)^2(x^2-1)^2(y^2-1)^2}{(x-y)^8(x+y)^8} \\ &\hspace{10mm}\times\left[\left(x^2-y^2+\frac{x^2+y^2-6}{2}\log\left|\frac{y^2-3}{x^2-3}\right|\right)^2 + \frac{\pi^2}{4}(x^2+y^2-6)^2\theta(y-\sqrt{3}) \right] \,.
\end{align}

We show in Fig.~\ref{fig:sgwb} the SGWB spectra corresponding to the scalar power spectra shown in Fig.~\ref{fig:PK}, compared to the sensitivity of future experiments. We also plot in cyan the recent claim of a detection from the NANOGrav collaboration~\cite{NANOGrav:2020bcs}. However, since there is no evidence for the so-called Hellings and Downs curve, which would be the smoking gun for a gravitational wave background detection, we interpret the data point as an upper limit (see~\cite{Zhao:2022kvz} for a recent study on the implications on peaked power spectra from inflation).

We note that $\Omega_{\rm GW}$ depends on the model parameters qualitatively in the same way as the curvature power spectrum. This is obvious because the integral in Eq.~\eqref{eq:sgwb} suggests that $\Omega_{\rm GW}(f)\propto\mathcal{P}_\mathcal{R}^2(f)$. We see that the background produced in Hybrid inflationary model typically falls in the sensitivity region of PTA experiments (magenta shaded area) and space-based gravitational waves interferometers such as LISA (turquoise-green shaded area) and more futuristic ones like BBO and DECIGO (grey shaded area). Some values of the parameter space, in particular those featuring a too-large amplification of the spectrum are already excluded by current PTA limits from the NANOGrav collaboration. Interestingly, the models excluded are those for which our analysis is less reliable and we should start to take into account stochastic effects the backreaction on the background dynamics during inflation,   see a discussion in Section \ref{sec:single}, and in the discussion of Fig. \ref{fig:PertEvo}.

Reconstructing the spectral shape of the SGWB is a target of future gravitational wave experiments~\cite{Caprini:2019pxz,Flauger:2020qyi}, it is thus important to provide simple templates for $\Omega_{\rm GW}(f)$ that can be readily used in such searches.  Since the computation of the primordial curvature power spectrum, and therefore of the induced SGWB, has to be performed numerically, such template is intended to be purely phenomenological and a one-to-one connection with the model parameters is not straightforward. Nevertheless, having a template that closely matches the exact result has the clear advantage of simplicity. In case of a detection of the phenomenological template, the exact signal can be compared numerically and constraints on the true model parameters can be set.

Focusing for simplicity on LISA, we see that, depending on where the peak is located, we may need a more complex template to capture the signal. For example, if the peak falls outside the sensitivity, the SGWB can be just described by a simple power law, as for the blue line in the top-left panel of Fig.~\ref{fig:sgwb}. If, on the other hand, the peak is exactly in the sensitivity range of LISA, a simple power-law would not be precise enough to describe the richer structure of $\Omega_{\rm GW}(f)$ and some more advanced functions to describe the peak will be needed. This is the case of the green line in the top-right panel of Fig.~\ref{fig:sgwb}.

As we can see, a simple lognormal shape for $\Omega_{\rm GW}$, i.e. a Gaussian in log-space, captures very well the region close to the peak, but it fails to describe the power-law behaviors at frequencies smaller and larger than the peak frequency $f_{\rm peak}$. While this accuracy is probably acceptable for LISA, whose sensitivity band is narrow enough not to see this deviation, it would be nice to have an even more accurate template, suitable for the next-generation of space based experiments. Also, this template is important to describe situations, such as the green line in the bottom-left panel of Fig.~\ref{fig:sgwb}, where the background falls in the sensitivity range of multiple experiments.

We see that a simple broken power law is not able to reproduce our signal either. A  solution is to smooth the transition between the two power-laws at the pivot frequency $f_{\rm peak}$ and adopt the following smoothed broken power-law template:

\begin{equation}
	h^2\Omega_{\rm GW}(f)=A\,\frac{\mathcal{F}(f,\,f_{\rm peak},\,n_1,\,n_2,\,\Delta)}{\mathcal{F}(f_{\rm peak},\,f_{\rm peak},\,n_1,\,n_2,\,\Delta)} \ ,
\end{equation}
where
\begin{equation}
	\mathcal{F}(f,\,f_{\rm peak},\,n_1,\,n_2,\,\Delta)=\frac{\left[\frac{f}{f_{\rm peak}}\left(-\frac{n_1}{n_2}\right)^{\Delta}\right]^{n_1}   }{
\left\{\frac{1}{2}	\left[1+\left(\frac{f}{f_{\rm peak}}\left(-\frac{n_1}{n_2}\right)^{\Delta}\right)^{1/\Delta}\right] \right\}^{\Delta(n_1-n_2)}} \ .
\end{equation}
As can be seen, with a proper choice of parameters, the template accurately matches the exact result and can therefore used in phenomenological searches for the SGWB produced in our model and similar ones. Indeed, we note that the GW signal produced by our model is also common to other realizations of hybrid inflation, that however produce different predictions at large scales, as previously discussed. The synergy of large and small scale data would be instrumental in pinning down the fundamental embedding of the hybrid inflation mechanism.

\section{Hybrid polynomial attractors}
\label{sec:polynomial}
The hybrid attractor scenario described in this paper was mainly devoted to the so-called exponential  attractors, such as $V_0 \tanh^2 \big ({\varphi \over \sqrt{6\alpha} } \big)$, where the potential of the field $\vp$ approaches the plateau exponentially,
 \be \label{exp}
V(\vp) = V_{0} \left(1-e^{-\vp/\mu}+...\right)  \ .
\ee 
But there is yet another broad class of cosmological attractors, where the potential approaches the plateau  more slowly, not exponentially but as inverse powers of the inflaton field, 
 \be \label{pol}
 V \sim V_{0}\left(1 -{\mu^{k}\over \vp^{k}}+... \right) \ ,
 \ee
where $k$ can be any (integer or not) positive constant. The simplest examples of such potentials are given by
\be 
V = V_0\,   { \varphi^{k} \over \varphi^{k}+  \alpha^{k}}\, , 
\label{Vcan4}\ee
They were called polynomial attractors~\cite{Kallosh:2022feu}.  Such potentials may appear in several different contexts, such as KKLTI inflation~\cite{Kachru:2003sx}, pole inflation~\cite{Kallosh:2019hzo}, and also as a special version of $\alpha$-attractors~\cite{Kallosh:2022feu}.

In all of these cases, at  $\alpha\ll 1$ in the large $N$ limit these potentials have universal attractor predictions, independently of other  details of the potential indicated by the ellipsis $...$ in Eq.~\rf{pol}.
In particular, the spectral index $n_{s}$ depends only on $k$~\cite{Kallosh:2018zsi}:
\be\label{genns}
n_{s} = 1-{2\over  N }{k+1\over k+2} \ .
 \ee

Here we will consider the hybrid inflation scenario where instead of the potential ${m^{2}\over 2} \phi^{2} $ we will use a potential $V = {m^{2}\over 2}  {\alpha^{2} \varphi^{2} \over \varphi^{2}+  \alpha^{2}} $. At $\vp^{2} \ll \alpha^{2}$ this potential is given by the familiar expression  ${m^{2}\over 2} \vp^{2} $, but at $\vp^{2} \gg \alpha^{2}$ it approaches a plateau $V_{0} = {m^{2}\alpha^{2}\over 2}$. 

Ignoring, for simplicity, the hybrid inflation uplifting, these models have universal predictions~\cite{Kallosh:2018zsi}
\be\label{genns2}
n_{s,{\rm KKLTI}} = 1-{3\over 2N }  \ , \qquad r = {\sqrt 2 \alpha\over N^{3/2}} \ .
 \ee

The corresponding hybrid inflation potential based on this particular version of the polynomial attractor is given by 
\be
\label{eq:H-poly2}
V_{\rm {\rm poly}}(\chi,\varphi) =M^{2}\left[{({\chi^2 -  \chi_{0}^{2}})^2\over 4\chi_{0}^{2}} 
+{\tilde m^{2}\over 2}   {\alpha^{2} \varphi^{2} \over \varphi^{2}+  \alpha^{2}}   +\frac{\tilde g^2}{2}  \vp^{2}\chi^2 +d\chi\right].
\ee
 
	\begin{figure}
		\begin{center}
				\includegraphics[width=.475\columnwidth]{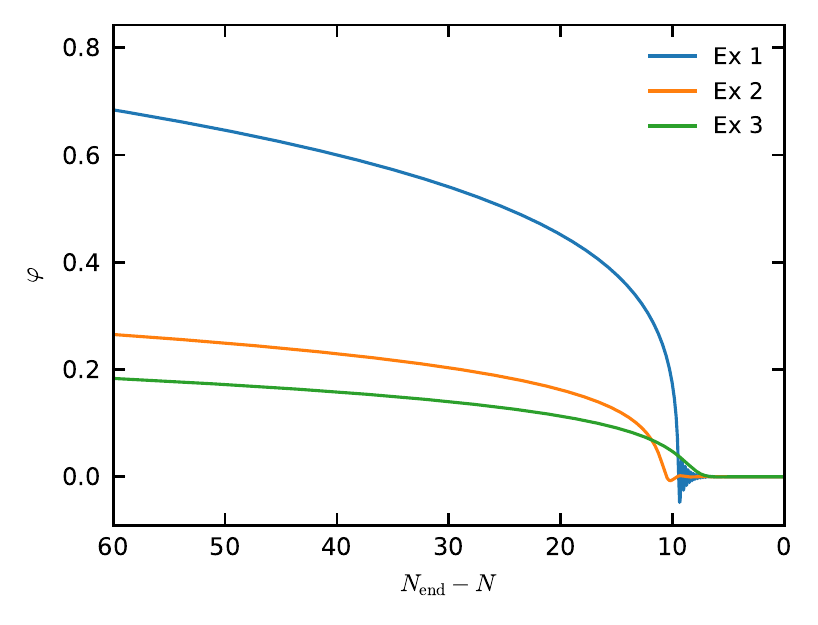}	
\includegraphics[width=.5\columnwidth]{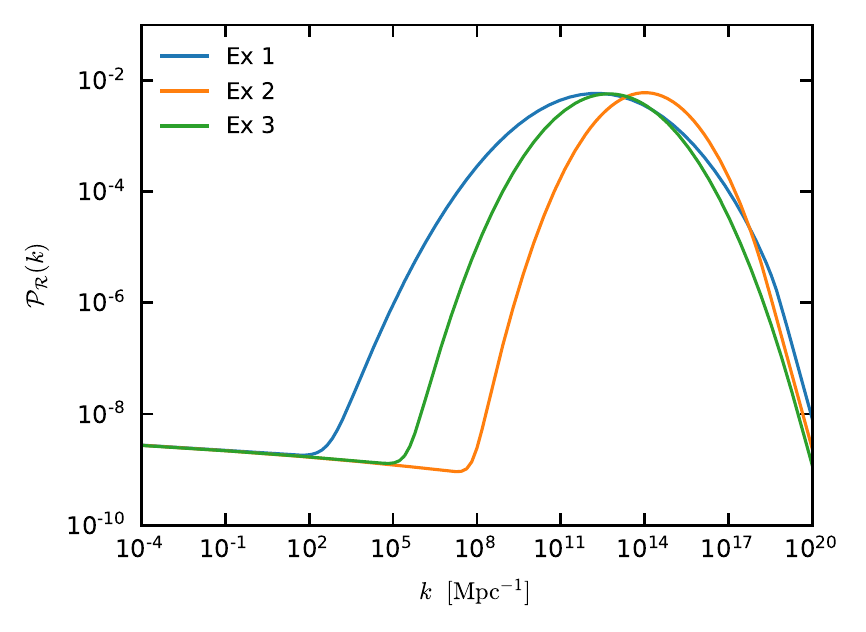}	
			\end{center}
		\caption{\footnotesize	\label{fig:KKLTI}   [Left] Evolution of the field $\varphi$ in the hybrid polynomial attractors model. [Right] Primordial power spectra. The parameters for each example are reported in the Table in the main text. }  
	\end{figure}
		 In Fig.~\ref{fig:KKLTI}, we provide some examples of the evolution of the field $\varphi$ in this model, together with the associated power spectra.	We fix the following parameters	$\alpha=0.05$, $\lambda =1/\chi_0^2=0.3$ and $M$ is chosen to match the COBE normalization, as usual. The small value of $\alpha$ is chosen here to take advantage of the attractor regime of polynomial attractors~\cite{Kallosh:2019hzo}.  The other parameters are varied according the following table. As discussed in the previous Section, the number of $e$-folds that is used in Eq.~\eqref{genns2} is $N_{\rm KKLTI}=N_*-\Delta N =55-\Delta N$, where we assume $N_*=55$.
	\begin{center}\begin{tabular}{|l|l|l|l|l|l|l|l|}
			\hline  
			&    $d$ & $\tilde{m}\alpha$ & $\tilde{g}$& $\Delta N$&$n_{s,{\rm KKLTI}}$&$n_{s,\,{\rm num}}$ \\\hline
			Example 1                 & $-2\times10^{-7}$ & $0.7$  &$2$&$9.43$&$0.9671$ & $0.9670$ \\         \hline
			Example 2                 & $-10^{-7}$ & $0.3$  &$6$&$11.00$&$0.9659$ & $0.9652$ \\         \hline
			Example 3                 & $-6\times10^{-8}$ & $0.05$  &$8$&$8.91$&$0.9674$ & $0.9662$ \\         \hline
	\end{tabular}\end{center}
	
	We can see that the main features of the peak in the power spectrum are essentially the same as in the exponential hybrid attractor models. The main difference is that the predictions at large scales follow a different attractor. Like in the previous Section, we observe that increasing $\tilde{m}\alpha$ brings the analytical predictions into closer agreement with the numerical results for $n_s$.   However, unlike in the case of exponential attractors, here this allows us to take full advantage of the attractor nature of the model. Indeed, since now $n_s=1 -3/2(N_*-\Delta N)$, the attractor predictions are consistent with Planck data even for the value of $\Delta N\sim 10$ needed to get a phenomenologically interesting peak at small scales in this model. Therefore in this context we do not need to rely on uplifting to shift the attractor prediction upwards to larger values of $n_{s}$. However, we can use this mechanism if we want to consider models with larger values of $\Delta N$, or if we want to increase $n_{s}$ even further.

\section{Discussion} 
\label{sec:conclusions}

 Models of inflation producing a large amplitude peak at small scales are the subject of a very active field of research, primarily because they potentially lead to the production of PBHs and a SGWB testable with future gravitational wave facilities. Historically, hybrid inflation~\cite{Linde:1991km,Linde:1993cn} was one of the first models where such an amplification of power was considered~\cite{Garcia-Bellido:1996mdl}, see also \cite{Randall:1995dj,Kawasaki:1997ju,Yokoyama:1998pt,Frampton:2010sw,Kawasaki:2016pql,Choi:2021yxz,Spanos:2021hpk,Kawai:2022emp}. In this paper, we investigated the possibility to generate perturbations with a very high peak in the spectrum in  its  recently developed  $\alpha$-attractor generalization~\cite{Kallosh:2022ggf} of the hybrid inflation scenario   and  discussed  its consequences for PBH and SGWB production. 

This scenario describes an inflaton field $\varphi$ interacting with a Higgs-type field $\chi$ in such a way that the latter becomes tachyonic when $\varphi$ becomes smaller than some critical value $\varphi_{c}$.  In the simplest cases, the absolute value of the tachyonic mass of the field $\chi$ becomes much greater than the Hubble constant $H$  at $\varphi <\varphi_{c}$. Is such cases, the field $\chi$ falls down to  the minimum of  the potential at $\chi = \chi_{0}$ and inflation abruptly ends at   $\varphi  \approx \varphi_{c}$   \cite{Linde:1991km,Linde:1993cn}. 

However, for $\chi_{0} \gg 1$   the tachyonic mass of the field $\chi$ at $\varphi  \lesssim \varphi_{c}$  remains  much smaller  than $H$.  In such cases,   inflation continues while the field $\chi$ slowly rolls down  towards the minimum of $V(\chi)$. As we explained in this paper, in this case, at $\varphi < \varphi_{c}$, the universe enters the eternal inflation regime: In some parts of the universe inflation eternally continues at $\chi = 0$, whereas in many other parts inflation ends within a relatively small number of $e$-foldings $\Delta N$. This leads to enormous perturbations of density on the scale proportional to $e^{\Delta N}$, which results in copious production of PBHs and eternally inflating parts of the universe, including eternally inflating topological defects.

This is a very general property of this class of models, which does not require any fine-tuning, so it is very easy to produce a  very high peak of perturbations and PBH in this scenario. The only problem is how to keep the situation under control, to avoid overproduction of  both PBHs and eternally inflating topological defects inside the observable part of the universe.

One way to do it is to reduce $\chi_{0}$, which stops eternal inflation and  suppresses the amplitude  of perturbations \cite{Garcia-Bellido:1996mdl,Clesse:2015wea}. However, this is not sufficient to solve the problem of supermassive topological defects. Another problem is how to make a large bump at some intermediate range of $e$-foldings and simultaneously have perturbation with $n_{s} \sim 0.96 - 0.97$ compatible with the Planck data.

In this paper, we  showed  that it is possible to solve all of these problems in the $\alpha$-attractor generalization of hybrid inflation~\cite{Kallosh:2022ggf} if we add a tiny linear term $\mu^{3} \chi$  to the potential. This term pushes the inflationary trajectory slightly off the ridge of the potential. This solves the problem of topological defects and simultaneously decreases the amplitude of the perturbation in a controllable way. The position and the width of the peak can be controlled by other parameters of the hybrid inflation potential.

 	Phenomenologically, key features of this model are the agreement with  CMB measurements from Planck and the production of PBHs and SGWB at smaller scales, that offer nice prospects for testing it in the future. Indeed, many realizations of hybrid inflation were excluded by the earliest Planck release~\cite{Planck:2013jfk}.
 Thanks to the $\alpha$-attractor modification of the model, one can find a broad range of parameters where $n_{s}$ and $r$ are consistent with the latest Planck/BICEP/Keck Array results. 	
 On smaller scales, the model is flexible enough to produce bumps at different locations and with different shapes, making it possible to produce very light to heavy PBHs, possibly constituting the totality of the Cold Dark Matter in our universe if the peak is located at the right position. While the formation of PBHs is mainly sensitive to the overall amplitude of the peak, future observations of the induced SGWB have the potential to unveil its spectral shape over  orders of magnitude in frequency. Our model thus constitute a primary target of future searches for stochastic backgrounds of gravitational waves. To this purpose, we have also derived a very simple template that perfectly captures the frequency profile of the gravitational wave energy spectrum, and can be readily used by the gravitational wave community to test our model. 
  We stress that the GW signal described by our template, first proposed in our paper, is quite generic to the hybrid inflation mechanism. However, the embedding of hybrid inflation into the $\alpha$-attractor framework leads to predictions at large scales that are consistent with current data, making the synergy of GW experiments with large scale structure measurements crucial to identify the origin of the hybrid inflation mechanism.

 We note that  the hybrid attractor model proposed in~\cite{Kallosh:2022ggf}, and further developed  in this paper, represents a  simple generalization of the basic hybrid inflation model   \cite{Linde:1991km,Linde:1993cn}. In the Appendix \ref{sec:sugra} it is shown how this scenario can be implemented in supergravity. One can also generalize this two-field model to include many other fields. In particular, instead of the inflaton potential ${m^{2}\over 2} \varphi^{2}$ one can consider O(2) invariant potential  ${m^{2}\over 2} (\varphi_{1}^{2}+\varphi_{2}^{2})$. In such model instead of domain walls we would have global strings. If we consider a potential ${m^{2}\over 2} (\varphi_{1}^{2}+\varphi_{2}^{2}+\varphi_{3}^{2})$ we would have global monopoles. By considering more general potentials for the field $\chi$, including  hilltop inflationary potentials  such as $(\chi^{4}-\chi_{0}^{4})^{2}$, or the Coleman-Weinberg potential used in new inflation, one may have a second inflationary stage with an extremely small value of $\chi_{0}$. In such models one may find a different way to control the amplitude of the spectrum and avoid problems with topological defects. In fact, as shown in \cite{Vilenkin:2018zol,Matsuda:2005ey}, in certain cases such defects may independently contribute to the PBH production.

\section*{Acknowledgement}
R.K. and A.L. are  supported by SITP and 
by the US National Science Foundation Grant PHY-2014215. 
A.L. is grateful to Juan Garcia-Bellido for many useful discussions and collaboration on  related projects. F.F. acknowledges financial support by the agreement n. 
2020-9-HH.0 ASI-UniRM2 ``Partecipazione italiana alla fase A della missione LiteBIRD".

\appendix

\section{Supergravity implementation of hybrid attractors}\label{sec:sugra}

 \subsection{Hybrid  $\alpha$-attractors with a canonical waterfall field, adding axions}
In the context of the PBH production hybrid inflation was studied  in supersymmetric F- and D-term inflation \cite{Clesse:2015wea}.
All fields in these models have canonical kinetic terms, therefore the choice of parameters consistent with CMB as well as leading to PBH production is not easy to make. For example in D-term supergravity model the Planck-like FI term and a superpotential with large coupling are required.

Our models have plateau potentials with non-canonical geometric inflaton at the first stage of inflation, which facilitates their agreement with the CMB data.

The $\alpha$-attractor versions of hybrid inflation models were proposed and studied in  \cite{Kallosh:2022ggf}. The first stage of inflation was defined by a plateau potential for the field $\varphi$ originating from a geometric hyperbolic disk geometry with the \K curvature ${\cal R} = -{2\over 3\alpha}$. The waterfall phase of hybrid inflation is due to a second field $\chi$ which was also defined in  \cite{Kallosh:2022ggf} by a geometric hyperbolic disk geometry with the \K curvature ${\cal R} = -{2\over 3\beta}$. 

In the models we study here we take the waterfall $\chi$ to be a canonical field, $\beta \to \infty$. For our purpose  to study high peaks in inflationary perturbations any value of $\beta>1$  is suitable, not much depends on it. Small values of $\beta$ are more interesting if we design the second stage of inflation as a quintessence cosmology, which is not the purpose of this paper. Therefore, in our case a canonical waterfall field is the simplest possibility. 

Here we will  present the supergravity version of the cosmological models  for a geometric inflaton $\varphi$ and a canonical waterfall field $\chi$
 studied in this paper. These models can also be obtained from 
 supergravity versions of general hybrid inflation in \cite{Kallosh:2022ggf} in the limit $\beta \to \infty$.

We start as follows
 \be 
{ {\cal L} \over \sqrt{-g}} =  {R\over 2}  -  {(\partial_{\mu} \phi)^2\over 2\bigl(1-{\phi^{2}\over 6\alpha}\bigr)^{2}} -  {(\partial_{\mu} \chi)^2\over 2} - V(\chi,\phi)   \,  ,
\label{cosmoAA}\ee 
where
\be
 V(\phi,\chi) =M^2  \Big( V_{\rm hybrid}(\chi,\phi) + V_{\rm lin}(\chi)\Big)\, , 
\label{pot}\ee
Here
 \be\label{hybridA}
V_{\rm hybrid} (\phi,\chi) =
 {m^2\over
2 }\phi^2 + {g^2\over 2    }\phi^2\chi^2  +{({\chi^2 -  \chi_{0}^{2}})^2\over 4\chi_{0}^{2}}  \ ,
\ee
\be
V_{\rm lin}(\chi)=  d  \chi \ . \ee
To be careful, one should also subtract a tiny constant $\approx d\,\chi_{0}$ to ensure smallness of the cosmological constant at the minimum, but this term can be ignored in the investigation of inflation.

Upon   transformation to canonical variables  $
\phi = \sqrt {6 \alpha}\, \tanh{\varphi\over\sqrt {6 \alpha}}$  we find
 \be 
{ {\cal L} \over \sqrt{-g}} =  {R\over 2}  - {1\over 2} (\partial_{\mu} \varphi)^2 -    {1\over 2}(\partial_{\mu} \chi)^2 - V(\varphi, \chi)   \,  .
\label{model}\ee 
The hybrid  potential \rf{pot} as a function of $\vp$ and $\chi$ becomes  \rf{hybridab}
\bea\label{hybridPot}
V(\varphi, \chi) =M^2 \Big[
3 \alpha \, \tanh^{2}{\varphi\over\sqrt {6 \alpha}} (\tilde m^{2 } + \tilde g^2 \,  \chi^2 ) +
  {({\chi^2 -  \chi_{0}^{2}})^2\over 4\chi_{0}^{2}} + d \chi  \Big] \ .
\eea
The model given in eqs. \rf{model}, \rf{hybridPot} is a model with 2 bosonic fields. In the main part of the paper we studied the cosmological properties of this model.

We would like to embed this bosonic model into a supergravity model. Such an embedding is not unique, we choose here a relatively simple one. 
The first step is to relate each of the scalars $\vp$ and $\chi$ to  complex scalars $Z_1, Z_2$. This means also that two scalars of the original bosonic models  are supplemented by two axions.
\be\label{ZZ}
Z_1= z e^{i\theta_1}
 = \tanh{\vp\over \sqrt{6\alpha}} e^{i\theta_1} \ , \qquad Z_2 = {1\over \sqrt 2}\chi e^{i\theta_2} \ .
\ee
Now we need to  make sure that there is a minimum in $\theta_i$ directions  at $\theta_i=0$, so that both axions are stabilized at their vanishing value. 

Many cosmological supergravity models associated with string theory and uplifting anti-D3 branes which were proposed and studied during the last few years involve an additional nilpotent chiral superfield  $X$ such that $X^2=0$. It represents non-linearly realized supersymmetry of the Volkov-Akulov type. 

The reason why the embedding proposed in \rf{ZZ} is relatively simple is the fact, established earlier in \cite{Carrasco:2015uma,Carrasco:2015pla,Kallosh:2017wnt},
that there is a universal geometric mechanism of stabilization of the inflaton partner during inflation (or in general, of the imaginary part of the complex scalar), based on the bisectional curvature. In our case, this is a statement that the fields ${\rm Im} \, Z_i$ can be stabilized at ${\rm Im}\,  Z_i=0$, i. e. at $\theta_1=\theta_2=0$, which reduces our supergravity model to the bosonic cosmological model we study in this paper.
In presence of the nilpotent superfield one can add to the \K potential a bisectional curvature term of the form 
\be
K_{\rm bisec}= X\bar X A(Z, \bar Z) {\rm Im}^2 Z
\label{bi}\ee
for each of the axions ${\rm Im}\,  Z$ which we want to stabilize. These terms with appropriate choice of the function $A(Z, \bar Z)$ add a positive mass terms to each of the axions at all positions in inflationary trajectory.

In fact, very often our earlier models in \cite{Kallosh:2017wnt} were stabilized at the vanishing values of the axions even without the need for bisectional curvature terms. This feature is provided by special properties of these models which we also used in  \cite{Kallosh:2021vcf} and will use here.

\subsection{Supergravity model with stabilized axions}
There are 3 chiral superfields, $Z_1, Z_2$ and a nilpotent one $X$. The  \K potential and the superpotential  are:
\be
K=K_{X\bar X}   X \bar{X}-3 \alpha  \log \left(1-{Z_1} \overline{{Z_1}}\right)+{Z_2} \overline{{Z_2}}\, ,  \qquad K_{X\bar X} = \frac{F_X^2 }{ {F_X}^2+ {\cal V}\left({Z_1},\overline{{Z_1}},{Z_2},\overline{{Z_2}}\right)} \ ,
\label{K}\ee
\be
W=M (F_X X+W_0)  \left(1-Z_1^2\right)^{\frac{3 \alpha }{2}}  e^{-\frac{{Z_2}^2}{2}} \ .
\ee
Here ${\cal V}\left({Z_1},\overline{{Z_1}},{Z_2},\overline{{Z_2}}\right)$ is an arbitrary function. To match the potential of the hybrid $\alpha$-attractors \rf{hybridPot} we will take
\be
 {\cal V}=M^{2} \left\{ 3\alpha  {Z_1} \overline{Z_1} \Big( \tilde m^2 + \tilde g^2{Z_2}  \overline{Z_2}\Big )+{1\over 4 \chi_0^2} \left[\left(\frac{{Z_2}+\overline{{Z_2}}}{\sqrt{2}}\right)^2 - \chi_0^2\right]^2 + d   \frac{{Z_2}+\overline{{Z_2}}}{\sqrt{2}} \right\} \ .
\label{calV}\ee
The total potential 
\be
V^{\rm super}= e^K (|DW|^2- 3|W|^2)  
\label{total}\ee
as a function of complex fields $Z_1, Z_2$ at $X=0$ is complicated, but at real $Z_1, Z_2$ it is reduced to \rf{hybridPot} where in addition there is a cosmological constant term $\Lambda = F_X^2- 3W^2$
\be
V^{\rm super}|_{Z_i= \overline Z_i ,  X=0} =\Lambda + V(\vp,\chi) \ .
\ee

Using \K invariance of the potential \rf{total} we can present an alternative form of $K$ and $W$ which helps to explain the reason why the axions are easy to stabilize in this class of models. The first term in $K$ is the same as in \rf{K} but the second and the third, as well as $W$ are different due to a \K transformation preserving the potential
\be
K=K_{X\bar X}   X \bar{X}-{3 \alpha\over 2}   \log {\left(1-{Z_1} \overline{{Z_1}}\right)\over (1-Z_1^2) (1- \overline{{Z_1}}^2)}- 
{1\over 2} ({Z_2} -\overline{{Z_2}})^2 \ ,
\ee
\be
W=M\, (F_X X+W_0) \ .
\ee
The \K potential $-{3 \alpha\over 2}   \log {\left(1-{Z_1} \overline{{Z_1}}\right)\over (1-Z_1^2) (1- \overline{{Z_1}}^2)}- 
{1\over 2} ({Z_2} -\overline{{Z_2}})^2$ vanishes at $Z_i= \overline Z_i$ by design, and it has terms proportional to axions $-(Z_i- \overline Z_i)^2$. The shift symmetry for axions $Z\to Z + ic$ is therefore broken already by $K$ and in most cases axions are stabilized even without additional terms with bisectional curvature \rf{bi},  as shown in \cite{Kallosh:2017wnt}. 

It was also important here that one can  make a choice of the function ${\cal V}\left({Z_1},\overline{{Z_1}},{Z_2},\overline{{Z_2}}\right )$ in eq. \rf{K} in the \K potential of the nilpotent field $K_{X\bar X}$ supporting the stabilization of the axions initiated by the choice of the \K potential of the  chiral superfields $Z_i$.

In particular, for the parameters used in this paper for the model in \rf{hybridPot} we have checked that axions are stabilized at $Z_i= \overline Z_i$ and they have large positive masses even without bisectional curvature terms 
\rf{bi}. For more general parameters,  in any case the bisectional curvature terms \rf{bi} can always be added and therefore we have promoted our bosonic models to  supergravity models consistently.

\bibliographystyle{JHEP}
\bibliography{references}
\end{document}